\documentclass[11pt]{article}
\usepackage{graphicx}
\usepackage{epsfig}
\usepackage{array}
\usepackage[figuresright]{rotating}
\usepackage{amsmath}
\usepackage{amssymb}
\usepackage{amsfonts}
\usepackage{appendix}

\usepackage{cite}
\begin{document}
\begin{center}
{\LARGE{\bf Polarizabilities of the nucleon and
spin dependent photo-absorption}  
  }\\[1ex] 
Martin Schumacher\\mschuma3@gwdg.de\\
Zweites Physikalisches Institut der Universit\"at G\"ottingen,
Friedrich-Hund-Platz 1\\ D-37077 G\"ottingen, Germany
\end{center}

\begin{abstract}
The polarizabilities $\alpha$ (electric), $\beta$ (magnetic) and $\gamma_\pi$
(backward spin) of the nucleon are investigated in terms of the degrees of 
freedom of the
nucleon using recent results for the CGLN amplitudes and resonance couplings
$A_{1/2}$ and $A_{3/2}$. The photon
excitation strengths of the excited states are given in terms of partial
integrated photoabsorption cross sections and resonant contributions of the 
Gerasimov-Drell-Hearn (GDH) sum rule. As a test of the present
predictions,  cross section differences 
$(\sigma_{3/2}-\sigma_{1/2})$ of the excited states are compared with data
recently measured at MAMI (Mainz) and ELSA (Bonn). In order to explain
differences between proton and neutron, radiative widths of the
excited states are compared with nonrelativistic and relativistic predictions
based on the $SU(6)$ harmonic oscillator (HO) quark model. A complete list of
partial contributions from the $s$-channel and the $t$-channel are given
for the polarizabilities. 
\end{abstract}

\section{Introduction}

The present work is a continuation of a systematic series of studies 
\cite{schumacher06,schumacher07a,schumacher07b,schumacher08}
on the 
electromagnetic structure of the nucleon, following experimental work
on Compton scattering and  a comprehensive  review on this topic
\cite{schumacher05}. These recent  investigations  have shown
\cite{schumacher06,schumacher07a,schumacher07b,schumacher08}
that a  systematic study of all partial resonant and nonresonant
photo-excitation processes of the nucleon and of their relevance for the
fundamental structure constants of the nucleon 
as there are the electric
polarizability ($\alpha$), the magnetic polarizability ($\beta$) and the
backward spin-polarizability ($\gamma_\pi$)
is essential for
an understanding of the electromagnetic structure of the nucleon. 
In addition it has been found that
the structure of the constituent quarks and their
coupling 
to pseudoscalar and scalar mesons is  important for 
the understanding of the electric and 
magnetic polarizabilities 
and of the backward spin-polarizability. The reason is that
neutral scalar mesons couple to two photons with parallel planes
of linear polarization and, therefore, make  sizable  $t$-channel
contributions to the electric and magnetic polarizabilities  whereas neutral
pseudoscalar mesons couple to two photons with perpendicular planes
of linear polarization and, therefore,  make a sizable  $t$-channel
contribution to the backward spin-polarizability.   In the course of these
studies it has been found desirable to have a better understanding of the 
level structure of the nucleon and of the cross section components
entering into the resonant parts of the polarizabilities. Especially, it had to
be clarified  whether characteristic differences found between the excited
states of the proton and the neutron can  be related to the quark-structure
of these particles. 
Since the backward
spin-polarizability depends on the helicity dependent photo-absorption cross
section, also this cross section was investigated.

This latter part of the work was
made possible  by recent precise measurements of helicity dependent cross
sections carried out at MAMI (Mainz) and ELSA (Bonn)
using circularly polarized photons
and spin-polarized proton and deuteron targets\cite{GDH1,GDH2,GDH3,GDH4,GDH5,GDH6,GDH7,GDH8,GDH9,GDH10,GDH11,GDH12,GDH13,GDH14}.
These studies have led to important data which have been used
for a test of the Gerasimov-Drell-Hearn (GDH) sum rule. The GDH sum rule
relates  the energy-weighted difference $(\sigma_{3/2}-\sigma_{1/2})$
of spin dependent cross sections to the square of the anomalous magnetic
moment $\kappa$ of the nucleon and in this way makes an important contribution
to the studies of the spin-structure of the nucleon. All aspects of the
experiment and the interpretation of the data have been nicely reviewed 
in a recent article \cite{helbing06} (see also \cite{drechsel04}). 

A further motivation for the present work 
was provided by the fact that the knowledge of
CGLN amplitudes and of the resonance couplings $A_{1/2}$ and $A_{3/2}$
has been improved to a large extent due to recent very precise 
MAID (Mainz) analyses of photo-absorption data in terms of CGLN amplitudes
\cite{drechsel07} (see also \cite{dugger07}). 
This made it possible to considerably improve on the prediction of 
spin and multipolarity dependent photo-absorption cross sections and, therefore,
also on the polarizabilities of the nucleon.

\section{Outline of the method}

Walker \cite{walker69,karliner73} writes the integral cross sections for pion
photoproduction in terms of helicity ``elements'' (or amplitudes)
$A_{l\pm}$ and $B_{l\pm}$, given in the form
\begin{eqnarray}
&&\sigma_T=\frac12 (\sigma_{1/2}+\sigma_{3/2}),\nonumber\\
&&\sigma_{1/2}=\frac{8\pi q}{k}\sum_{l=0}^\infty (l+1)(|A_{l+}|^2+
|A_{(l+1)-}|^2),
\nonumber\\
&&\sigma_{3/2}=\frac{8\pi q}{k}\sum_{l=0}^\infty \frac14 [l(l+1)(l+2)]
(|B_{l+}|^2+|B_{(l+1)-}|^2),\label{walker}
\end{eqnarray}
where the subscript notation of the $A$'s and $B$'s is the same as that
of the CGLN; e.g. $B_{l\pm}$ refers to a state with pion orbital angular
momentum $l$ and total angular momentum $J=l\pm\frac12$. The $A$'s and $B$'s
differ in the absolute values of the helicities $\Lambda=|\lambda|$ of the 
initial states given by $\lambda=\lambda_\gamma-\lambda_p$, where
$\lambda_\gamma$ is the helicity of the incident photon and $\lambda_p$
the helicity of the nucleon in the initial state. For the $A$'s the helicity
is $\Lambda=1/2$ for the $B$'s $\Lambda=3/2$. The quantity $q$ is the 
3-momentum of the pion and $k$ the 3-momentum of the photon in the c.m. system.

Walker \cite{walker69}
finds the following relations  between the helicity elements and the CGLN
multipole coefficients:
\begin{eqnarray}
&& A_{k+}=\frac12[(k+2)E_{k+}+k M_{k+}],\quad B_{k+}=E_{k+}- M_{k+},
\nonumber\\
&&A_{(k+1)-}=\frac12[-kE_{(k+1)-}+(k+2)M_{(k+1)-}],\nonumber\\
&&B_{(k+1)-}=E_{(k+1)-}+M_{(k+1)-}. \label{multipoles}
\end{eqnarray} 
From isospin consideration it has been derived that the amplitudes for
meson photoproduction $A$ are composed of $A^{(1/2)}$ and $A^{(3/2)}$,
referring to final states of definite isospin $(\frac12)$ or 
$(\frac32)$. Furthermore, there is an amplitude $A^{(0)}$ 
corresponding to the isoscalar transition. This latter amplitude
makes a contribution to $I=1/2$ only. Therefore, the amplitudes 
\begin{eqnarray}
&&_pA^{(1/2)}=A^{(0)}+\frac13 A^{(1/2)}\nonumber\\
&&_nA^{(1/2)}=A^{(0)}-\frac13 A^{(1/2)}
\label{Aisospin-1}
\end{eqnarray}
may be introduced. Furthermore, with 
\begin{eqnarray}
&&A^{(+)}=\frac13 (A^{(1/2)}+ 2  A^{(3/2)}),\nonumber\\
&&A^{(-)}=\frac13 (A^{(1/2)}- A^{(3/2)}),
\label{Aisospin-2}
\end{eqnarray}
the physical amplitudes may be expressed by the isospin combinations 
(see, e.g., \cite{drechsel99,ericson88})
\begin{eqnarray}
&& A(\gamma p\to n \pi^+)=\sqrt{2}(A^{(-)}+A^{(0)})=\sqrt{2}
\left(_pA^{(1/2)}-\frac13 A^{(3/2)}\right),\\
&&A(\gamma p\to p \pi^0)=A^{(+)}+A^{(0)}=
_pA^{(1/2)}+\frac23 A^{(3/2)},\\
&&A(\gamma n\to p \pi^-)=-\sqrt{2}(A^{(-)}-A^{(0)})=\sqrt{2}
\left(_nA^{(1/2)}+\frac13 A^{(3/2)}\right),\\
&&A(\gamma n\to n \pi^0)=A^{(+)}-A^{(0)}=
-_nA^{(1/2)}+\frac23 A^{(3/2)}.
\end{eqnarray}
From these equations we arrive at the isospin decompositions
\begin{eqnarray}
&&|A|^2({\rm proton)}=|A(\gamma p \to n\pi^+)|^2+|A(\gamma p \to p \pi^0)|^2=
3|_pA^{(1/2)}|^2+\frac23|A^{(3/2)}|^2,\label{decomp-1}\\
&&|A|^2({\rm neutron)}=|A(\gamma n \to p\pi^-)|^2+|A(\gamma n \to n\pi^0)|^2=
3|_nA^{(1/2)}|^2+\frac23|A^{(3/2)}|^2 \label{decomp-2}.
\end{eqnarray}
Making use of (\ref{decomp-1}) and (\ref{decomp-2}) we 
arrive at the 
helicity and isospin dependent cross sections for the most important
multipoles. These relations are valid for the 1$\pi$ channel only.

\begin{eqnarray}
&&\sigma^{(1\pi)}_{1/2}(0+)=\frac{8 \pi
  q}{k}\left[3\left|_{(p.n)}E^{(1/2)}_{0+}\right|^2+\frac23\left|E^{(3/2)}_{0+}
  \right|^2\right], \label{13}\\
&& \sigma^{(1\pi)}_{3/2}(0+)=0, \label{14}\\
&&\sigma^{(1\pi)}_{1/2}(1-)=\frac{8 \pi
  q}{k}\left[3\left|_{(p.n)}M^{(1/2)}_{1-}\right|^2+\frac23\left|M^{(3/2)}_{1-}
  \right|^2\right], \label{15}\\
&& \sigma^{(1\pi)}_{3/2}(1-)=0, \label{16}\\
&&\sigma^{(1\pi)}_{1/2}(1+)=\frac{8 \pi
  q}{k}\frac12\left[3\left|3_{(p.n)}E^{(1/2)}_{1+}+_{(p.n)}M^{(1/2)}_{1+}\right|^2+\frac23\left|3E_{1+}^{(3/2)}+
  M^{(3/2)}_{1+} \right|^2\right], \label{17}\\
&&\sigma^{(1\pi)}_{3/2}(1+)=\frac{8 \pi q}{k}
  \frac32\left[3\left|_{(p,n)}E^{(1/2)}_{1+}-_{(p.n)}
M^{(1/2)}_{1+}\right|^2+\frac23\left|E^{(3/2)}_{1+}-
M^{(3/2)}_{1+}\right|^2\right],
  \label{18}\\
&&\sigma^{(1\pi)}_{1/2}(2-)=\frac{8 \pi
  q}{k}\frac12 \left[3\left|-_{(p.n)}E^{(1/2)}_{2-}+3_{(p.n)}M^{(1/2)}_{2-}\right|^2+\frac23\left|-E_{2-}^{(3/2)}+
 3 M^{(3/2)}_{2-} \right|^2\right],\label{19}\\
&&\sigma^{(1\pi)}_{3/2}(2-)=\frac{8\pi
  q}{k}\frac32\left[3\left|_{(p,n)}E^{(1/2)}_{2-}+_{(p.n)}M^{(1/2)}_{2-}\right|^2+\frac23\left|E^{(3/2)}_{2-}+M^{(3/2)}_{2-}\right|^2\right],\label{20}\\
&&\sigma^{(1\pi)}_{1/2}(2+)=\frac{8\pi
  q}{k}3\left[3\left|2\,_{(p,n)}E^{(1/2)}_{2+}+_{(p.n)}M^{(1/2)}_{2+}\right|^2+\frac23\left|2\,E^{(3/2)}_{2+}+M^{(3/2)}_{2+}\right|^2\right],\label{21}\\
&&\sigma^{(1\pi)}_{3/2}(2+)=\frac{8\pi
  q}{k}6\left[3\left|_{(p,n)}E^{(1/2)}_{2+}-_{(p.n)}M^{(1/2)}_{2+}\right|^2+\frac23\left|E^{(3/2)}_{2+}-M^{(3/2)}_{2+}\right|^2\right],\label{22}\\
&&\sigma^{(1\pi)}_{1/2}(3-)=\frac{8\pi
  q}{k}3\left[3\left|-\,_{(p,n)}E^{(1/2)}_{3-}+2\,_{(p.n)}M^{(1/2)}_{3-}\right|^2+\frac23\left|-\,E^{(3/2)}_{3-}+2\,M^{(3/2)}_{3-}\right|^2\right],\label{23}\\
&&\sigma^{(1\pi)}_{3/2}(3-)=\frac{8\pi
  q}{k}6\left[3\left|_{(p,n)}E^{(1/2)}_{3-}+_{(p.n)}M^{(1/2)}_{3-}\right|^2+\frac23\left|E^{(3/2)}_{3-}+M^{(3/2)}_{3-}\right|^2\right],\label{24}\\
&&\sigma^{(1\pi)}_{1/2}(3+)=\frac{8\pi
  q}{k}\left[3\left|5\,_{(p,n)}E^{(1/2)}_{3+}+3\,_{(p.n)}M^{(1/2)}_{3+}\right|^2+\frac23\left|5\,E^{(3/2)}_{3+}+3\,M^{(3/2)}_{3+}\right|^2\right],\label{25}\\
&&\sigma^{(1\pi)}_{3/2}(3+)=\frac{8\pi
  q}{k}15\left[3\left|_{(p,n)}E^{(1/2)}_{3+}-_{(p.n)}M^{(1/2)}_{3+}\right|^2+\frac23\left|E^{(3/2)}_{3+}-M^{(3/2)}_{3+}\right|^2\right].\label{26}
\end{eqnarray}

It is generally assumed \cite{drechsel07,drechsel99a}
that the inelasticity ($2\pi$ and $\eta$)   
correction is independent of the helicity\footnote{The same branching
  correction $\Gamma/\Gamma_{\pi}$ is used to relate $A_{1/2}$ and $A_{3/2}$
to the CGLN amplitudes.}.
This  implies
that the difference of helicity dependent cross sections 
$(\sigma_{3/2}-\sigma_{1/2})$
and the total photo-absorption cross section 
$\sigma_T=\frac12 (\sigma_{3/2} +\sigma_{1/2})$ are proportional to each
other:
\begin{equation}
\sigma_{3/2}-\sigma_{1/2}=A_n\,\,\frac12 (\sigma_{3/2} +\sigma_{1/2})
\label{proportional}
\end{equation}
where for non-mixed multipolarity we have $A_n=-2$ for $E_{0+}$ and $M_{1-}$,
$A_n=+1$ for $M_{1+}$ and $E_{2-}$, $A_n=+2/3$ for $M_{2+}$ and $E_{3-}$,
and $A_n=1/2$ for $M_{3+}$. For mixed multipolarity 
the proportionality constant is given by 
\begin{equation}
A_n=\left[ \frac{2(\sigma_{3/2}- \sigma_{1/2})}{\sigma_{3/2}+\sigma_{1/2}}
\right]_{\rm res.}
\label{proportionalityconstant}
\end{equation}
 and can easily be calculated from the  CGLN amplitudes
\cite{drechsel07} at resonance maximum where
only the imaginary parts are of importance. However,  thereafter
slight corrections
are possible because of a slight energy dependence of $A_n$ 
which may be determined by considering the CGLN amplitudes
over the  entire resonance or by adjusting to experimental data.
The latter procedure will be carried out in subsection 4.2.

The advantage of using Eq. (\ref{proportional}) is that the very convenient
Walker parameterization \cite{walker69,armstrong72}
of nucleon resonant cross sections derived for
$\sigma_T=\frac12 (\sigma_{3/2} +\sigma_{1/2})$ can also be applied to 
 $(\sigma_{3/2}-\sigma_{1/2})$. For convenience we give the Walker
 parameterization in the following (see \cite{armstrong72} for the presently
used definitions and the choice of damping factors):
\begin{eqnarray}
&&I=I_r\left(\frac{k_r}{k}\right)^2\frac{W_r^2\,\Gamma\,\Gamma^*_\gamma}{
(W^2-W^2_r)^2+W^2_r\Gamma^2},\label{28}\\
&&\Gamma=\Gamma_r\left(\frac{q}{q_r}\right)^{2l+1}\left(\frac{
q^2_r+X^2}{q^2+X^2}\right)^l,\label{29}\\
&&\Gamma^*_\gamma=\Gamma_r\left(\frac{k}{k_r}\right)^{2\,j_\gamma}
\left(\frac{k^2_r+X^2}{k^2+ X^2} \right)^{j_\gamma},\label{29a}\\
&&s=2\,\omega\, m+m^2, \label{30}\\
&&\omega= \mbox{ photon energy in the lab. system},\label{31}\\
&&m= \mbox{ nucleon mass },\label{31a}\\
&&W^2=s,\label{31a}\\
&&k=|{\bf k}|=\frac{s-m^2}{2\sqrt{s}},\label{32} \\
&&|{\bf k}|=\mbox{photon 3-momentum in the c.m. system},\label{33}\\
&&q=|{\bf q}|=\sqrt{E^2_\pi-m^2_\pi};\quad E_\pi=\frac{s-m^2+m^2_\pi}{2\sqrt{s}},
\label{34}\\
&&|{\bf q}|= \mbox{$\pi$ 3-momentum in the c.m. system},\label{35}\\
&&j_\gamma,\,\, \mbox{multipole angular momentum of the photon,}\label{36}\\
&&l, \,\,\mbox{single $\pi$ angular momentum.}\label{37}
\end{eqnarray}
The quantities $I_r, \Gamma_r,k_r,q_r$ are the peak cross section,
the width of the resonance, the photon 3-momentum and pion 3-momentum at
resonance in the c.m. system.
The damping constants $X$ are $X=160$ MeV for the $P_{33}(1232)$-resonance and
$X=350$ MeV else.
The peak cross-section $I_r$ introduced in (\ref{28}) is given by\footnote{
Please note that the quantity $\Gamma^*_\gamma$ in Eq. (\ref{29a}) has a
different definition as the  quantity $\Gamma_\gamma$ in the following Eqs. 
(\ref{peak})  and (\ref{Gg}).}
\begin{equation}
I_r=2\,\pi\,\frac{1}{k^2_r}\frac{2\,J+1}{2\,J_0+1}\frac{\Gamma_\gamma}{\Gamma},
\label{peak}
\end{equation}
where $J$ and $J_0$ are the spins of the excited states and the ground state, 
respectively, $\Gamma_\gamma$ the photon width and $\Gamma$ the  total
width of the resonance. The photon width $\Gamma_\gamma$ may be expressed 
through the resonance couplings $A_{1/2}$ and $A_{3/2}$ by the relation
\cite{PDG} \begin{equation}
\Gamma_\gamma=\frac{k^2_r}{\pi}\frac{2M_N}{(2J+1)M_R}[|A_{1/2}|^2+
  |A_{3/2}|^2],
\label{Gg}
\end{equation}
where $M_N$ and $M_R$ are the nucleon and resonant masses. Combining
(\ref{peak}) and (\ref{Gg}) we arrive at 
\begin{equation}
I_r=\frac{2M_N}{M_R\Gamma}[|A_{1/2}|^2+|A_{3/2}|^2].
\label{Irfinal}
\end{equation}
One  advantage of the present approach over other approaches based on the 
CGLN amplitudes is that the quantity $I_r$ defined in (\ref{Irfinal})
contains the branching correction $\Gamma/\Gamma_\pi$.

It is of interest to compare the present approach with previous approaches.
At the resonance energy the partial cross sections given in   
(\ref{13}) -- (\ref{26}) are related to the resonance couplings introduced by
Arndt et al. \cite{arndt90} through the relations (see Eq. (\ref{Irfinal}))
\begin{equation}
\sigma^{(1\pi)}_{1/2,\,3/2}(l\pm)(\Gamma/\Gamma_\pi)=\frac{4\,M_N}{M_R\,
\Gamma}|A^{1/2,\,3/2}_{l\pm}|^2
\label{arndt}
\end{equation}
where $\Gamma_\pi$ is the $\pi N$ elastic width. These resonance couplings
are related to the electric and magnetic multipoles via the relations

\begin{eqnarray}
&& A^{1/2}_{l+}=-\frac12 [(l+2)\,\overline{E}_{l+}+l\,\overline{M}_{l+}],\label{44}\\
&& A^{3/2}_{l+}=\frac12\sqrt{l(l+2)}[\overline{E}_{l+}-\overline{M}_{l+}],\label{45}\\
&& A^{1/2}_{(l+1)-}=-\frac12[l\,\overline{E}_{(l+1)-}-(l+2)\overline{M}_{(l+1)-}],
\label{46}\\
&& A^{3/2}_{(l+1)-}=-\frac12\sqrt{l(l+2)}[\overline{E}_{(l+1)-}
+\overline{M}_{(l+1)-}],\label{47}
\end{eqnarray}
where the above barred multipoles are related to the CGLN $(E,M)$
multipoles at resonance energy $W_r$ through the relations
\begin{equation}
(\overline{E},\overline{M})=C \left[ \frac{(2j+1)\pi q_r W_r \Gamma^2_r}
{k_r M_N \Gamma_\pi} \right]^{1/2}(E,M).
\label{48}
\end{equation}
The factor $C$ in Eq. (\ref{48}) is  $\sqrt{2/3}$ for isospin $\frac32$
and $-\sqrt{3}$ for isospin $\frac12$. It is easy to see that the approach  
outlined in Eqs. (\ref{arndt}) -- (\ref{48}) leads to the same result as the
relations (\ref{13}) -- (\ref{26}) at the resonance energy $W_r$.

\section{The level structure of the nucleon}

A comprehensive outline of the structure of baryons in terms of quark models
has recently been given by the Particle Data Group \cite{PDG}. Therefore,
in this section we may restrict the discussion to those aspects which
are of special interest for the present investigation and not covered in that
article.

It has been successful \cite{PDG,moorhouse66,faiman68,lichtenberg69,copley69,feynman71,close79,koniuk80,capstick86,manley87,close90,giannini90,capstick92}
to classify nucleon resonances in terms of the 
 spin-flavor SU(6) harmonic oscillator (HO) model.
Empirically we know that the ground state of the harmonic oscillator $(N=0)$
is a $({\bf 56},L^P=0^+)$ super-multiplet state. 
The first excited state $(N=1)$ consists
of the $({\bf 70},1^-)$ super multiplet only. 
The second excited state ($N=2$) consists of the 
$({\bf 56},0^+)$, $({\bf 56},2^+)$, $({\bf 70},0^+)$ and $({\bf 70},2^+)$
super multiplets of which $({\bf 70},0^+)$ and $({\bf 70},2^+)$ are barely
seen experimentally. Reasons for the preference of the 
$({\bf 56},0^+)$, $({\bf 56},2^+)$ super multiplets have been discussed
e.g. in \cite{close79,manley87,faiman68,lichtenberg69}.

\subsection{Empirical level scheme based on the $SU(6)$-HO quark model}
\begin{table}[h]
\caption{Nucleon resonant states and meson photoproduction for the first three
oscillator shells.
The integrated cross sections in column 4 ($I_{int.}$)
are based on the resonance couplings $A_{1/2}$ and $A_{3/2}$ of the
proton (p) and the neutron (n). Within the subgroups the states are ordered
with respect to the size of the total angular momentum $J$.}
\vspace{3mm}
\setlength{\extrarowheight}{5pt}
\begin{tabular}{llcc}
\hline
\hline
& $^{2S+1}L^{(I)}_J$&& $I_{int.}$\\
&&&                    [$10^3\,\mu$b MeV]\\
\hline
\hline
$N=0$, $L=0$ & $^2S^{(1/2)}_{1/2}$\,$(8,2)$ & $P_{11}(939)$& p\quad n\\
$({\bf 56},0^+)$             
& $^4S^{(3/2)}_{3/2}$\,$(10,4)$ & $M1,(E2)\to P_{33}(1232) \to M^{(3/2)}_{1+},
(E^{(3/2)}_{1+})    $ & $80.0 \quad 80.0$\\
\hline
\hline
$N=2$, $L=0$ & $^2S^{(1/2)}_{1/2}$\,$(8,2)$ & $M1 \to P_{11}(1440) 
\to\, _{p,n}M^{(1/2)}_{1-}$&
$3.4\quad 1.3$ \\
$({\bf 56},0^+)$           & $^4S^{(3/2)}_{3/2}$\,$(10,4)$
& $M1,(E2) \to P_{33}(1600) 
\to M^{(3/2)}_{1+},(E^{(3/2)}_{1+})$& $0.7\quad 0.7$\\
\hline
\hline
$N=1$, $L=1$& $^2P^{(1/2)}_{1/2}$\,$(8,2)$
& $E1 \to S_{11}(1535) 
\to\, _{p,n}E^{(1/2)}_{0+}$&
$8.8\quad 2.4$\\
$({\bf 70},1^-)$    
& $^2P^{(1/2)}_{3/2}$\,$(8,2)$
& $E1,(M2) \to D_{13}(1520) \to\, _{p,n}E^{(1/2)}_{2-},\,
(_{p,n}M^{(1/2)}_{2-})$& $34.6\quad 27.9$\\
\hline
$N=1$, $L=1$&    $^2P^{(3/2)}_{1/2}$\,$(10,2)$
& $E1 \to S_{31}(1620) \to E^{(3/2)}_{0+}$
&$0.9\quad  0.9$\\
$({\bf 70},1^-)$
&   $^2P^{(3/2)}_{3/2}$\,$(10,2)$
& $E1,(M2) \to D_{33}(1700) \to E^{(3/2)}_{2-},
\,(M^{(3/2)}_{2-})$&
$18.0\quad 18.0$\\
\hline
$N=1$, $L=1$& $^4 P^{(1/2)}_{1/2}$\,$(8,4)$
& $E1\to S_{11}(1650)
\to\,_{p,n} E^{(1/2)}_{0+}
$& $3.1\quad  0.2$\\
$({\bf 70},1^-)$          
& $^4P^{(1/2)}_{3/2}$\,$(8,4)$
& $E1,(M2)\to D_{13}(1700)\to\,
_{p,n}E^{(1/2)}_{2-},\, (_{p,n}M^{(1/2)}_{2-}) $& $0.4\quad  0.1$\\
          & $^4P^{(1/2)}_{5/2}$\,$(8,4)$
& $M2,(E3)\to   D_{15}(1675)
\to\,_{p,n}M^{(1/2)}_{2+},\, (_{p,n}E^{(1/2)}_{2+})  $& $0.7\quad  6.2$\\
\hline
\hline
$N=2$, $L=2$& $^2D^{(1/2)}_{3/2}$\,$(8,2)$
& $M1,(E2) \to P_{13}(1720) \to\, _{p,n}
M^{(1/2)}_{1+},\,
(_{p,n}E^{(1/2)}_{1+})$& $0.8\quad  1.0$\\
$({\bf 56},2^+)$
&$^2D^{(1/2)}_{5/2}$\,$(8,2)$  
& $E2,(M3) \to F_{15}(1680) \to\, _{p,n}E^{(1/2)}_{3-},\,
(_{p,n}M^{(1/2)}_{3-})$& $19.9\quad  2.2$\\
\hline
$N=2$, $L=2$& $^4D^{(3/2)}_{1/2}$\,$(10,4)$
& $M1\to P_{31}(1910) \to M^{(3/2)}_{1-}$
&$< 0.1 \quad < 0.1$\\
$({\bf 56},2^+)$  
&$^4D^{(3/2)}_{3/2}$\,$(10,4)$  & $M1,(E2) \to P_{33}(1920) \to
M^{(3/2)}_{1+},(E^{(3/2)}_{1+})$& $2.3\quad  2.3$\\
&$^4D^{(3/2)}_{5/2}$\,$(10,4)$ 
& $E2,(M3) \to F_{35}(1905) \to
E^{(3/2)}_{3-},(M^{(3/2)}_{3-})$& $2.7\quad  2.7$\\
&$^4D^{(3/2)}_{7/2}$\,$(10,4)$
& $M3,(E4) \to F_{37}(1950) \to
M^{(3/2)}_{3+},(E^{(3/2)}_{3+})$&$15.8\quad  15.8$\\
\hline
\hline
\end{tabular}
\label{singleparticletable-1}
\end{table}

The excitation spectrum of the nucleon constructed on the basis
of the classification scheme outlined   above is shown in Table
\ref{singleparticletable-1} and in the  level scheme of Figure 
\ref{secondshell} given in the appendix.
Column 1 of Table  \ref{singleparticletable-1} shows the oscillator 
quantum number $N$, the angular momentum $L$ of the HO subshell 
and the dimension of the corresponding super multiplet.
Column 2 shows the spectroscopic  configuration 
\begin{equation}
^{2S+1}L^{(I)}_J
\label{spectroscon}
\end{equation}
and the 
(SU(3),SU(2)) subgroup dimensions of states possible for the  given
harmonic oscillator subshell. In (\ref{spectroscon}) $S,L,J$ and $I$
are the spin, orbital angular momentum, total angular momentum and 
isospin of the state, where $L=0,1,2,\cdots$ has been represented by
$S,P,D.\cdots$, respectively. 
Within the subgroups the order of levels is according to the size of the
total angular momentum $J$.
Column 3 shows nucleon resonances and the
corresponding  electromagnetic and CGLN multipoles which have to be 
attributed to
the  spectroscopic  configurations of column 2. All the states
expected in the first three oscillator shells have been included,
except for the $N=2\,\,\, ({\bf 70},0^+)$ resonances $P_{11}(1710)$
which has the $***$ status. The 
*** $D_{35}(1930)$ resonance which energetically fits into the third oscillator
shell is a member of the  fourth oscillator shell because of its parity.
Column 4 shows the integrated
photo-absorption cross sections ($I_{int.}$) 
corresponding to the resonances and are given for the proton (p) and the
neutron (n). These numbers 
are described in more detail in appendix B.

In the 
second oscillator shell we find 7 levels in total. They belong to the
subgroups $(8,2)$ or  ($I=1/2$, $S=1/2$), $(8,4)$ or ($I=1/2$, $S=3/2$)
and $(10,4)$ or ($I=3/2$, $S=3/2$). The two states $D_{13}(1520)$ and
$S_{11}(1535)$ form the second resonance region of the nucleon photo-excitation
spectrum whereas the $D_{33}(1700)$ state makes a contribution to the third
resonance region in addition to the $F_{15}(1680)$, stemming from the third
oscillator shell. The other states contain less strength and do not lead to
pronounced structures in the nucleon photo-excitation spectrum. 
There are  8 nucleon levels of the third
oscillator shell. The $N=2$ $({\bf 56},0^+)$ state contains the Roper resonance
$P_{11}(1440)$ and the $\Delta$ resonance state $P_{33}(1600)$ build
upon the Roper resonance.  These two states do not lead to pronounced
structures in the photo-excitation spectrum of the nucleon. In contrast to this
the $N=2$ $({\bf 56},2^+)$ state contains two states, $F_{15}(1680)$ and the 
$F_{37}(1950)$,  which are the most prominent members of the third and fourth
resonance regions of the nucleon photo-excitation spectrum.

Table \ref{singleparticletable-1} reveals that within the subgroups
corresponding to a given set of quantum numbers $(N,L,I,S)$ the fully stretched
member, i.e. the one with the largest $J$, in general contains the largest
photo-excitation strength. States with $I=3/2$ have the same photo-excitation
strength for the proton and the neutron because the transition from the ground
state to the excited state is possible through an isovector transition
only. On the other hand, for states with $I=1/2$ the excitation is possible
via isoscalar and isovector transitions and, therefore, the proton and
neutron may contain different photo-excitation strengths. As a general rule we
find that for the proton the photo-excitation strength is larger than for the
neutron with the remarkable exception of the $D_{15}(1675)$ resonance where
the neutron contains a ten times larger photon excitation strength than 
the proton. This latter finding is in agreement with the 
Moorhouse selection  rule \cite{moorhouse66,faiman68}
according to which in the quark model the photon  width 
$\Gamma_\gamma(D^+_{15}(1675)\to p \gamma)=0$ whereas 
$\Gamma_\gamma(D^0_{15}(1675)\to n \gamma)\neq 0$. Further information
concerning the photon  widths of the levels of the nucleon obtained from
the quark model are given in the following subsection.

\subsection{Radiative widths $\Gamma_\gamma$ predicted in the quark model
\label{quark}}

Radiative  widths $\Gamma_\gamma$ of nucleon resonances 
in the nonrelativistic $SU(6)$-HO model
have been
derived by Faiman and Hendry \cite{faiman68}. In these calculations the 
oscillator spring constant $\alpha^2$ and the constituent-quark g-factor 
are treated as adjustable parameters. In the same model the resonance
couplings $A_{1/2}$ and $A_{3/2}$ are calculated by Copley et al. 
\cite{copley69}. In this work the g-factor g=1 is used but again
the oscillator spring constant is teated as an adjustable parameter.
The number given in that work, {\it viz.}  $\alpha^2=0.17$ GeV$^2$, together
with $g=1$ are used in general in later work. A complete list 
of results for the resonance couplings $A_{1/2}$ and $A_{3/2}$ are calculated
by Koniuk and Isgur \cite{koniuk80}.

The general structure of the expression describing the radiative width in the
nonrelativistic $SU(6)$-HO quark model is given by
\begin{equation}
\Gamma_\gamma=a\,k^3(k^2/\alpha^2)^N[g^2+b(\alpha^2/k^2)g+c(\alpha^4/k^4)]B
\label{general}
\end{equation}
\begin{table}[h]
\caption{Experimental nucleon radiative widths for the first three
oscillator shells compared with $SU(6)$-HO model predictions. (a)
nonrelativistic quark model\cite{faiman68,copley69,koniuk80}, (b)
relativized quark model  \cite{close90},
(c) relativized quark model \cite{capstick92}. 
}
\begin{center}
\vspace{3mm}
\setlength{\extrarowheight}{3pt}
\begin{tabular}{lllllllll}
\hline
&proton&&& $\Gamma_\gamma$ [MeV]&neutron&&&\\
\hline
& experiment & th.(a) & th. (b) & th.(c) & experiment& th.(a) & th. (b) 
& th.(c)\\
\hline
$ P_{33}(1232) 
$ & $0.69\pm 0.02$ &$0.33$& $0.41$&$0.37$
& $0.69\pm 0.02$&$0.33$&$0.41$& $0.37$\\
\hline
$P_{11}(1440) $& $0.15\pm 0.02  $& $0.02$&$0.004$& $0.0006$&
$   0.06\pm 0.03$ & $0.009$&$0.004$& $0.0013$\\
$P_{33}(1600) $& $0.02^{+0.05}_{-0.01}$  &$0.03$&$0.02$& $0.09$&  
$ 0.02^{+0.05}_{-0.01}$&$0.03$&$0.02$&  $0.09$\\
\hline
$S_{11}(1535) $&$0.36\pm 0.16   $&$1.3$&$1.2$& $0.26$
&$    0.09^{+0.14}_{-0.08}$&$0.7$&$0.5$& $0.18$ \\
$D_{13}(1520) $& $0.61\pm 0.04 $ &$0.35$&$0.40 $& $0.39$
& $   0.49 \pm 0.07$ &$0.38$&$0.51$& $0.31$\\
$S_{31}(1620)$ &$0.04\pm 0.03$ &$0.2$&$0.6$&  $0.4$
&$  0.04\pm 0.03$&$0.2$&$0.6$& $0.4$\\
$D_{33}(1700) $&$0.55\pm 0.15 $&$0.58$&$0.33$& $0.35$
&$ 0.55\pm 0.15$&$0.58$&$0.33$& $0.35$\\
$S_{11}(1650)$& $0.16\pm 0.10$  &$0.0$&$0.04$& $0.16$
& $   0.013^{+0.060}_{-0.010}$&$0.04$&$0.006$& $0.07$\\
$D_{13}(1700)$& $0.010^{+0.015}_{-0.010} $ &$0.0$&$0.09$& $0.03$
&$   0.03^{+0.09}_{-0.03}$ &$0.13$&$0.04$& $0.04$\\
$D_{15}(1675) $& $0.011\pm 0.008 $&$0.0$&$0.0$& $0.0003$
&$  0.10\pm 0.04$&$0.08$&$0.13$& $0.08$\\
\hline
$P_{13}(1720) $& $0.02^{+0.09}_{-0.02}$&$0.4$&$0.3 $& $0.03$
&$  0.03^{+0.09}_{-0.02}$&$0.03$&$0.009  $& $0.004$\\
$F_{15}(1680) $& $0.36\pm 0.06 $&$0.11$&$0.19 $& $0.09$
&$  0.04\pm 0.02$&$0.02$&$0.03 $& $0.018$\\
$P_{31}(1910) $&$0.001^{+0.007}_{-0.001} $&$0.03$&$0.03 $& $0.005$
&$  0.001^{+0.007}_{-0.001}$&$0.03$&$0.03$& $0.005$\\
$P_{33}(1920) $& $0.09\pm 0.06$&$0.07$&$0.04$& $0.015$
&$  0.09\pm 0.06 $&$0.07$&$0.04$&  $0.015$\\
$F_{35}(1905) $& $0.07\pm0.05 $&$0.09$&$0.08 $& $0.02$
&$  0.07\pm 0.05$&$0.09$&$0.08$& $0.02$\\
$F_{37}(1950) $&$0.33\pm 0.06 $&$0.07$&$0.14 $& $0.06$
&$  0.33\pm 0.06$&$0.07$&$0.14$& $0.06$ \\
\hline
\end{tabular}
\label{gammawidths}
\end{center}
\end{table} 
where $k$ is the photon 3-momentum in the c.m. system, $N$ the oscillator
quantum number  and $B$ a factor
containing the quark level magnetic moment $\mu=\mu_p$,  the 
factor $\exp(-k^2/(3\alpha^2))$
characterizing the harmonic oscillator and a kinematical factor. 
The specific $SU(6)$ spin-flavor properties of the nuclear levels are contained
in the quantities $a$, $b$ and $c$ which are rational numbers and may be found
in \cite{faiman68} or derived from the results given in
\cite{copley69} and \cite{koniuk80}. As far as the quantities  
$a$ are nonzero we 
find that for $I=1/2$
these number are smaller for the neutron than for the proton. The
proton-neutron ratio of the quantity $a$ is $9/4$ for $P_{11}$, $P_{13}$
and $F_{15}$ states, and $9$ for $S_{11}$ and $D_{13}$ states.
This is in qualitative 
agreement with the observation that the integrated cross section given in
Table \ref{singleparticletable-1} 
are in general smaller for the neutron than for the
proton. This means that qualitatively the difference in photo-excitation
strength of proton and neutron may be understood in terms of the 
$SU(6)$-HO quark model. In some cases as e.g. for the $D_{13}(1520)$ state 
the different factors $a$ for the proton and the neutron are compensated 
by the expression in square brackets so that the photo-excitation strengths
are almost the same for  the two nucleons.

It was pointed out in other works (see
e.g. \cite{feynman71,capstick86,close90,capstick92} 
and references therein) that the nonrelativistic version of the SU(6)-HO
model should be supplemented by relativistic corrections. Of these we give
special attention to the two latest calculation carried out by 
Close and Li \cite{close90} and by Capstick \cite{capstick92}.
Numerical results for the photon widths are given in Table 
\ref{gammawidths}. The experimental data are calculated from the resonance 
couplings given in  Table
\ref{resonanceparametersproton} and \ref{resonanceparametersneutron}.
The predictions (a) are based on the nonrelativistic
$SU(6)$-HO
model with the parameter $\alpha^2=0.17$ GeV$^2$ and $g=1$ as derived by
Copley et al \cite{copley69} and also used in later work. 
The predictions (b) have been obtained \cite{close90}
by supplementing the nonrelativistic results by spin-orbit terms.
The predictions (c) have been obtained in a
quark model \cite{capstick92} where an electromagnetic transition operator is
used containing relativistic corrections, and relativized-quark-model
wave functions. 

A special  feature is that in the nonrelativistic 
$SU(6)$-HO quark model the proton radiative widths of the $S_{11}(1650)$,
$D_{13}(1700)$ and $D_{15}(1675)$ are equal to zero. For the $D_{15}(1675)$
state this feature is known as the Moorhouse selection rule 
\cite{moorhouse66} whereas for the  $S_{11}(1650)$ and $D_{13}(1700)$
states two other states, {\it viz.} $S_{11}(1535)$ and $D_{13}(1520)$
with the same quantum numbers exist  so that the physical states
are linear combinations. The corresponding expressions are given in 
\cite{faiman68}. Mixing angles are given in \cite{PDG}.
The relativized
predictions (b) and (c) in general  are in better agreement with the 
experimental data than the 
nonrelativistic predictions 
(a) though some apparent drawbacks are visible
in all three cases. 
As a summary we may state, that qualitatively the nonrelativistic and the 
relativistic $SU(6)$-HO quark models explain many features of the radiative
widths. Especially it is explained that for $I=1/2$
the radiative widths of the neutron
are in general smaller than those of  the proton, 
except for the effects of the Moorhouse selection rule.

It should be mentioned that very recently progress has been made by 
developing a hyper-central constituent quark model with a meson 
cloud \cite{chen07}. The predicted resonance couplings $A_{3/2}$ and
$A_{1/2}$ for the $P_{33}(1232)$ resonance are in a better agreement with the
experimental values with the pion cloud than without the pion cloud. 
Further results are
expected for the $S_{11}(1535)$ and the $D_{13}(1520)$ resonances.

\section{The Gerasimov-Drell-Hearn sum rule}

Using dispersion theory as outlined in \cite{lvov97} and 
\cite{schumacher05} we arrive at 
\begin{equation}
\frac{1}{\omega}{\rm Re}\,g_0(\omega)=
{\rm Re}\,g(\omega)=-\frac{\alpha_e}{2}\left(\frac{\kappa}{m} \right)^2
+\frac{\omega^2}{4\pi^2}{\cal P}\int^\infty_{\omega_0}\frac{d\omega'}{\omega'}
\frac{\Delta \sigma(\omega')}{\omega'^2-\omega^2},
\label{GDHintegral}
\end{equation}
where $\alpha_e=1/137.04$ and
\begin{equation}
\Delta\sigma(\omega) = {\sigma_{1/2}(\omega)-\sigma_{3/2}(\omega)}.
\label{delsig}
\end{equation}
This leads to
\begin{equation}
{\rm Re}\,g(\infty)=-\frac{\alpha_e}{2}\left(\frac{\kappa}{m} \right)^2
-\frac{1}{4\pi^2}\int^\infty_{\omega_0}\frac{d\omega'}{\omega'}
\Delta \sigma(\omega').
\label{GDHinfty}
\end{equation}
\clearpage
\newpage
\begin{table}[h]
\caption{Lines 3--13:
Nucleon resonances observed via spin independent or helicity dependent
measurements of the photo-absorption cross section of the nucleon. The very
weak resonances $P_{33}(1600)$, $S_{31}(1620)$, $D_{13}(1700)$,
$P_{13}(1720)$ and  $P_{31}(1910)$  have been omitted. The
quantity $A_n$ in the second column is the scaling factor of 
Eq. (\ref{proportional}) for the leading
multipole, $A_n$(proton) and $A_n$(neutron) the corresponding scaling factors
obtained from the CGLN amplitudes in the resonance maximum. The CGLN
data have been taken from \cite{drechsel07}.
The quantities $I_{\rm GDH}$(proton) and $I_{\rm GDH}$(neutron)
are the contributions of the respective resonance to the GDH integral.  
Lines 15--17: Nonresonant contributions
to the GDH integral. Line 19: Predictions of the Regge model for contributions 
to the GDH integral according to \cite{helbing06}.
}
\begin{center}
\begin{tabular}{lllclc}
resonance &$A_n$&$A_n$(proton)& $I_{\rm GDH}$(proton)& $A_n$(neutron)& 
$I_{\rm GDH}$(neutron)\\
         & && [$\mu$b]&& [$\mu$b]\\
\hline
\vspace{2mm}
$P_{33}(1232)$&$+1$& $+1.26$&      $+291\pm 15$&$+1.26$& $+291\pm 15$\\
\vspace{2mm}
$P_{11}(1440)$&$-2$& $-2.0$  &    $-11.6\pm 2.7$&$-2.0$& $-4.4\pm 2.3$   \\
\vspace{2mm}
$D_{13}(1520)$&$+1$&$ +1.91$&  $+77.5\pm 7.8$& $+1.23$ &  $+40.2\pm 6.5$\\
\vspace{2mm}
$S_{11}(1535)$&$-2$&$ -2.0 $   & $-22.6 \pm 9.0$ & $-2.0$&  $-6.1\pm 3.5$ \\
\vspace{2mm}
$S_{11}(1650)$&$-2$& $-2.0$  & $ -6.4\pm 1.9$& $-2.0$ & $-0.5\pm 0.5$  \\
\vspace{2mm}
$D_{15}(1675)$&$+2/3$&$+0.67$  & $ +0.4\pm 0.4$&$+0.67$& $+4.2\pm 2.1$  \\
\vspace{2mm}
$F_{15}(1680)$&$+2/3$&$+1.88$&  $+35.0\pm 3.5$&$+0.76$&  $+1.5\pm 0.6$\\
\vspace{2mm}
$D_{33}(1700)$&$+1$&$\pm 0.0$&  $0.0\pm 3.5$&$\pm 0.0$& $0.0\pm 3.5$ \\
\vspace{2mm}
$F_{35}(1905)$&$+2/3$&$+1.31$&  $+2.8\pm 1.6$&$+1.31$&$+2.8\pm 1.6$ \\
\vspace{2mm}
$P_{33}(1920)$&$+1$&$+1.0$&  $+1.7\pm 1.2$&$+1.0$&$+1.7\pm 1.2$ \\
\vspace{2mm}
$F_{37}(1950)$&$+1/2$&$+0.56$&  $+5.6\pm 1.7$&$+0.56$& $+5.6\pm 1.7$\\
\hline
\vspace{2mm}
&resonant& sum& $374\pm 20$&&$336\pm 18$\\
\hline
\vspace{2mm}
nonres$(E_{0+})$&&& $-147$&&$-184$\\
\vspace{2mm}
nonres$((M,E)^{(1/2)}_{1+})$&&& $+13$&&$+47$\\
\vspace{2mm}
nonres$(M^{(3/2)}_{1-})$&&& $-9$&&$-9$\\
\hline
\vspace{2mm}
&present & total& $231\pm 20$&&$190\pm 18$\\
\hline
\hline
&Regge & predicted &-15&&+41\\
final result&Regge &included&$216\pm20$&&$231\pm 18$\\
&&sum rule& 205 &&233\\
\hline
\hline
\end{tabular}
\end{center}
\label{GDHprotonneutron}
\end{table}
\clearpage
\newpage
\noindent
There have been  many arguments accumulated over the years that $g(\infty)=0$
(see e.g. \cite{helbing06}). Assuming that this no-subtraction hypothesis
is fulfilled  we arrive at the frequently cited GDH sum rule
\begin{equation}
\int^\infty_{\omega_0}\frac{d\omega}{\omega}\left[\sigma_{3/2}
(\omega)-\sigma_{1/2}(\omega)
\right] =  2\pi^2\alpha_e\left(\frac{\kappa}{m}\right)^2.
\label{GDHint}
\end{equation}
By combining Eqs.  (\ref{GDHintegral}) and 
 (\ref{GDHint}) the GDH dispersion relation may be written in the form  
\begin{equation}
{\rm Re}\,g_0(\omega)=\omega\,{\rm Re}\,g(\omega)=\frac{\omega}{4\pi^2}{\cal P}
\int^\infty_{\omega_0}\frac{\omega'\,\Delta\sigma(\omega')}{\omega'^2-\omega^2}
d\omega'.
\label{GDHconventional}
\end{equation}
For comparison see e.g. \cite{schumacher05,helbing06,drechsel03}.

\subsection{Contributions  of  resonant and nonresonant excitation
  processes  to the GDH integral}

Explicit expressions for   the helicity and isospin dependent 
cross sections are given in  Eqs. (\ref{13}) to (\ref{26}).
For the $E_{0+}$ multipole both isospin components make a contribution
to the nonresonant cross section and up to 500 MeV this cross section is
purely nonresonant. 
Resonant contributions from
the $E_{0+}$ multipole show up at higher energies in the $S_{11}(1535)$,
the $S_{31}(1620)$ and the $S_{11}(1650)$ resonances. Therefore, in order 
to obtain the nonresonant $E_{0+}$ cross section above 500 MeV 
we have to make an
extrapolation. This is possible with sufficient precision by adjusting the
Born approximation to the experimental data.

For the $M_{1-}$ multipole the $I=3/2$
component makes a contribution to the nonresonant cross section 
whereas the $I=1/2$
component is exhausted by the $P_{11}(1440)$ resonance. 
 For the $M_{1+}$ and $E_{1+}$ multipoles the $I=1/2$
component makes a contribution to the 
nonresonant cross section whereas the $I=3/2$
component is exhausted by the $P_{33}(1232)$ resonance. 
For the nonresonant cross sections we make explicit use of the expressions
given in Eqs. (\ref{13}) to (\ref{18}). This implies that we take into account
the $1\pi$ channel only. This certainly is correct at low energies where the
main contributions to the GDH integral are located.

For the resonant contributions we make use of the expression in Eq. 
(\ref{proportional}) which relates the difference  of the helicity dependent
cross sections to the total photo-absorption cross section. The advantage of
this procedure is that use can be made of the Walker parameterization
of the total photo-absorption cross section and of the predictions 
for the peak cross sections 
obtained from the resonance couplings (see Eqs. (\ref{28}) to 
(\ref{Irfinal})). Through this procedure the $2\pi$
and $\eta$ photoproduction channels  are implicitly  taken into account.
The proportionality constants $A_n$ can be easily determined in case there is
only one multipole either electric $E_{l\pm}$ or magnetic $M_{\pm}$.
This is shown in column 2 of Table \ref{GDHprotonneutron}. 
In case of multipolarity mixing
strong deviations from the quantities $A_n$ given in column 2 of Table 
\ref{GDHprotonneutron}
are observed. In this case the quantities $A_n$ can be predicted  by evaluating
the appropriate expression as given in (\ref{13}) to (\ref{26}) at the
resonance peak using CGLN amplitudes \cite{drechsel07}. 
The result of this procedure is given in  columns 3 and 5 of Table 
\ref{GDHprotonneutron}.
It is interesting to
note that the $2\pi$ decay component of the $D_{13}(1520)$ has been 
experimentally observed in the reactions 
$\vec{\gamma} \vec{p}\to n \pi^+ \pi^0$
and  $\vec{\gamma} \vec{p}\to p \pi^0 \pi^0$ \cite{GDH5,GDH11}.
These decay components are taken care of in the present  analysis
by using  the total cross section of the $D_{13}(1520)$ resonance
with the $2\pi$ decay component included.

\begin{table}[h]
\caption{Non-Regge contributions to the GDH integral. The data of 
Workman and  Arndt, Sandorfi et al. and Drechsel and Krein contain the 
two-pion correction calculated by Karliner. The data of Fix and Arenh\"ovel
and of the present work contain two-pion corrections developed in the
respective works. 
}
\begin{center}
\begin{tabular}{lll}
\hline
&            $I^{\rm p}_{\rm GDH}$&  $I^{\rm n}_{\rm GDH}$\\
\hline
Karliner\cite{karliner73}& $261$ & $183$\\
Workman, Arndt \cite{workman92} & $260$ & $192$ \\
Sandorfi et al. \cite{sandorfi94} & $289$ & $160$ \\
Drechsel, Krein \cite{drechsel98} & $261$ & $180$ \\
Fix, Arenh\"ovel \cite{fix05} & $216.58$ & $190.51$   \\
present work& $231\pm 20$ &$ 190\pm 18$\\
\hline
experiment \cite{helbing06,GDH8,GDH14}& $226\pm 13$ &$214 \pm 35$\\
\hline
\hline
\end{tabular}
\end{center}
\label{GDHcomparison}
\end{table}

Another nonresonant contribution  is expected at high energies from 
diffractive effects and can be calculated using Regge parameterizations. 
Estimates carried out in \cite{helbing06}  yields 
$I^{\rm Regge}_{\rm GDH}({\rm proton})=\,-15 \mu$b and 
$I^{\rm Regge}_{\rm GDH}({\rm neutron})=\,+41 \mu$b. Including these
diffractive (Regge) contributions into the calculation we arrive at the 
final results $I_{\rm GDH}({\rm proton})=(216\pm 20)\mu$b and 
$I_{\rm GDH}({\rm neutron})=(231\pm 18)\mu$b.
It is of interest to compare the present approach to the GDH sum rule
with previous approaches. This is carried out in Table \ref{GDHcomparison}.
There is a general
satisfactory agreement between the different predictions. Experimental
data on helicity dependent photo-absorption cross sections have been 
measured at MAMI (Mainz) and ELSA (Bonn) 
\cite{GDH1,GDH2,GDH3,GDH4,GDH5,GDH6,GDH7,GDH8,GDH9,GDH10,GDH11,GDH12,GDH13,GDH14}. 
The experimental data for the
proton given in Table \ref{GDHcomparison} have  been taken from
\begin{equation}
I^{^1H}_{\rm run}(2.9 {\rm\,\, GeV})= 226\pm 5_{\rm stat}\pm 12_{\rm syst} 
\,\,\mu{\rm b}
\label{exp1}
\end{equation}
as given in \cite{helbing06,GDH8}. The experimental data for neutron have been
taken from
\begin{equation}
I^{^2H}_{\rm run}(1.8 {\rm\,\, GeV})= 440\pm 21_{\rm stat}\pm 25_{\rm syst} 
\,\,\mu{\rm b}
\label{cita2}
\end{equation}
as given in \cite{GDH14} by subtracting the value for the proton as given in 
(\ref{exp1}).
Comparing the different predictions with the experimental data, it appears
that the results of the present work are in closest agreement.

\subsection{Predicted  difference of helicity dependent cross sections 
 compared with experimental data}

Recent measurements at MAMI (Mainz) and ELSA (Bonn) have led to very valuable
data for the helicity dependence of the total photo-absorption cross section
of the proton and
the neutron. In the following we will compare experimental data obtained for
the proton with the predictions of the present approach. We restrict the
discussion to the proton because of the higher precision of the experimental
data as compared to the neutron.

\begin{figure}[h]
\includegraphics[width=0.45\linewidth]{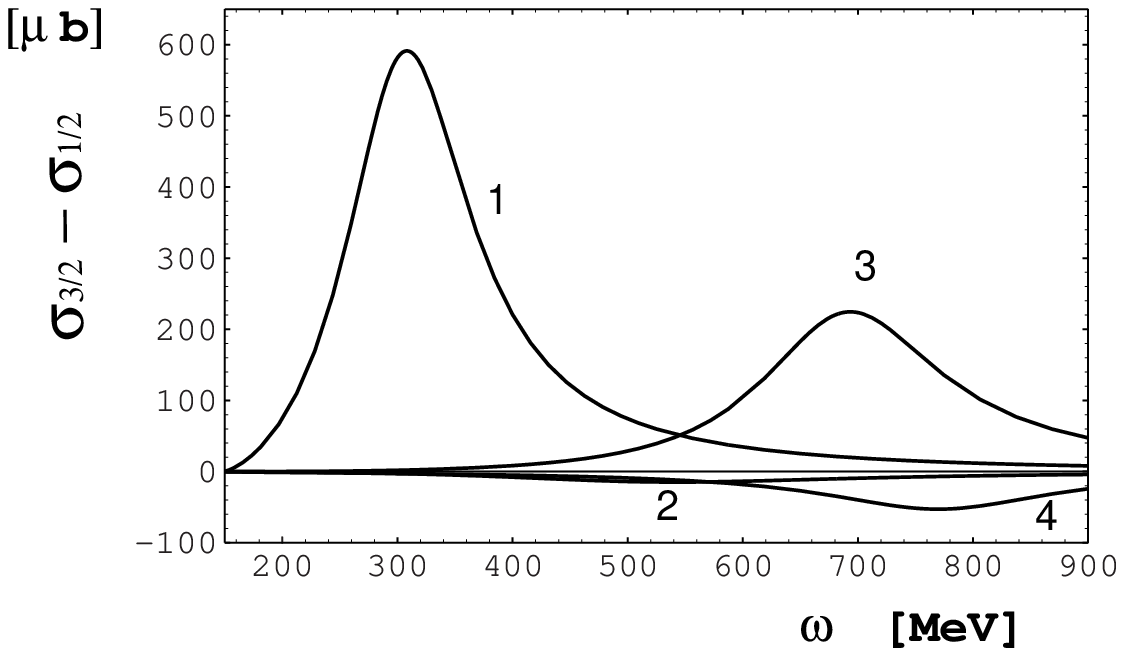}
\includegraphics[width=0.45\linewidth]{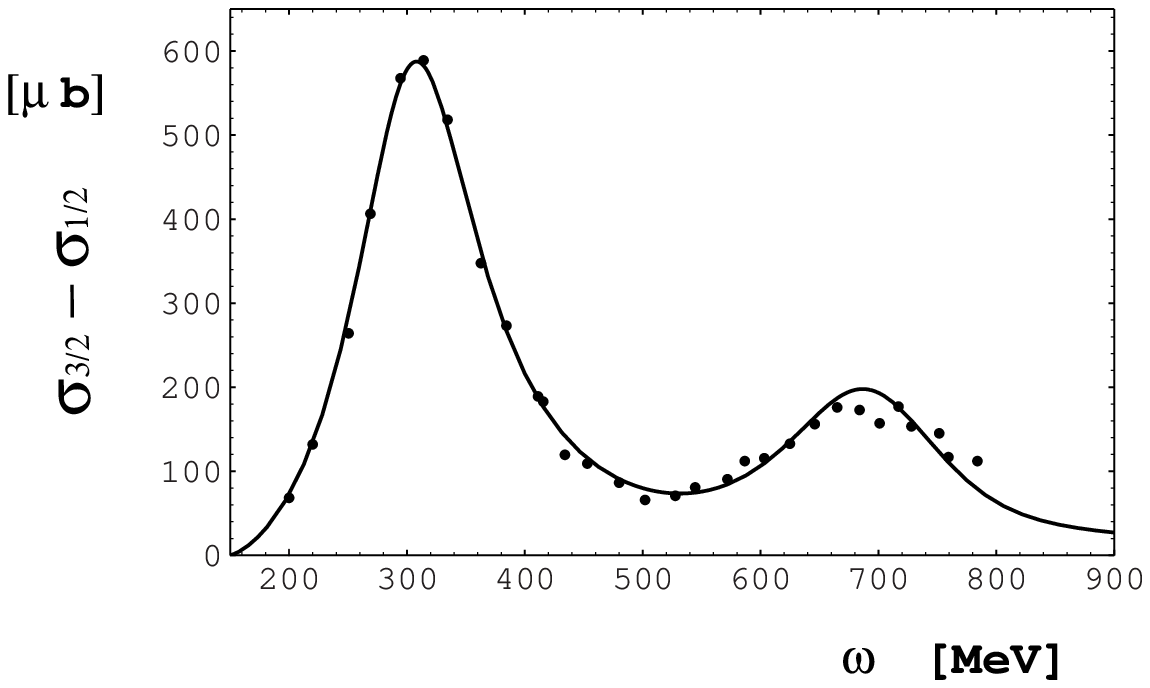}
\caption{Predicted  difference of helicity dependent 
cross sections for the proton
in the first and second resonance region compared with experimental data. 
The experimental data are taken from
\cite{GDH2}. The nonresonant contribution is eliminated from the figure by
a procedure described in the text. The contributions shown in the left panel
 are the resonances $P_{33}(1232)$ (1), $P_{11}(1440)$ (2),
$D_{13}(1520)$ (3), and $S_{11}(1535)$ (4).
}
\label{MAMI}
\end{figure}

In Figure  \ref{MAMI} we discuss experimental data obtained at MAMI (Mainz).
In order to clearly demonstrate the resonant structure of the 
first and the second resonance we have eliminated the  
nonresonant contribution from the figure and only keep the resonant
contribution. The baseline of the resonant contribution now is the abscissa.
Technically this has been achieved by adding the predicted nonresonant
contribution  to  the experimental data.     
The $P_{33}(1232)$ and
$D_{13}(1520)$ make strong positive contributions whereas the $P_{11}(1440)$
and the $S_{13}(1535)$ make small negative contributions. During the fitting
procedure it was found out that slight shifts of some of the parameters within
their margins of errors led to an improvement of the fit to the experimental
data. For the $P_{33}(1232)$ resonance these shifts are
1232 MeV $\to$ 1226 MeV for the position 
\begin{figure}[h]
\includegraphics[width=0.45\linewidth]{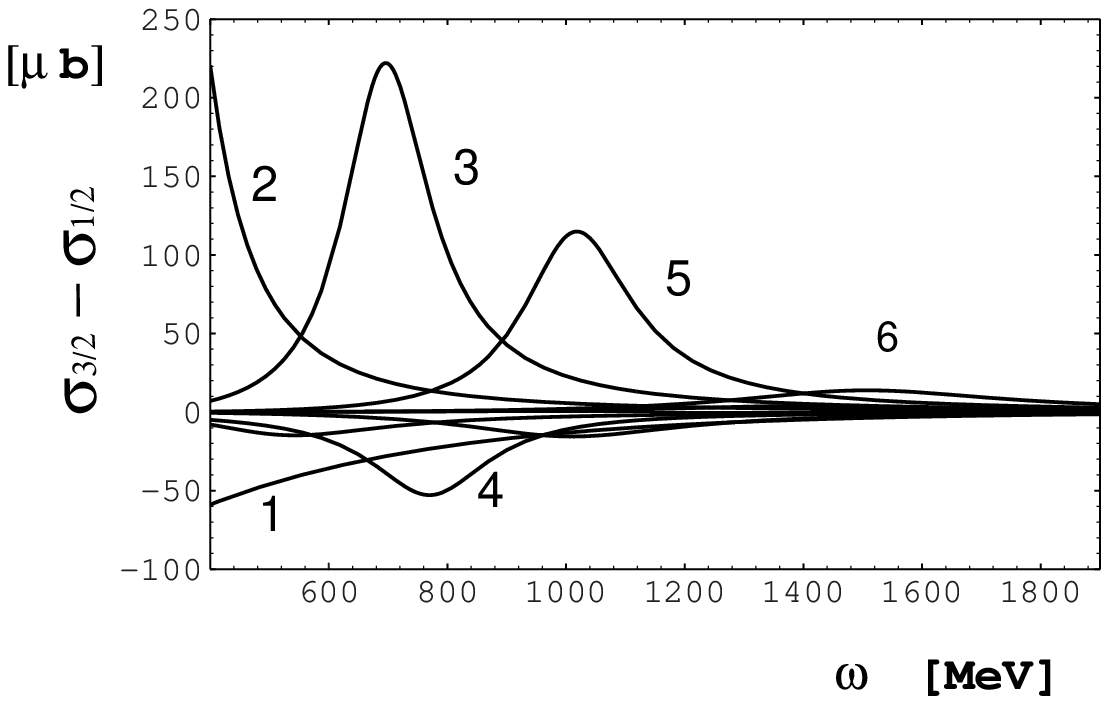}
\includegraphics[width=0.45\linewidth]{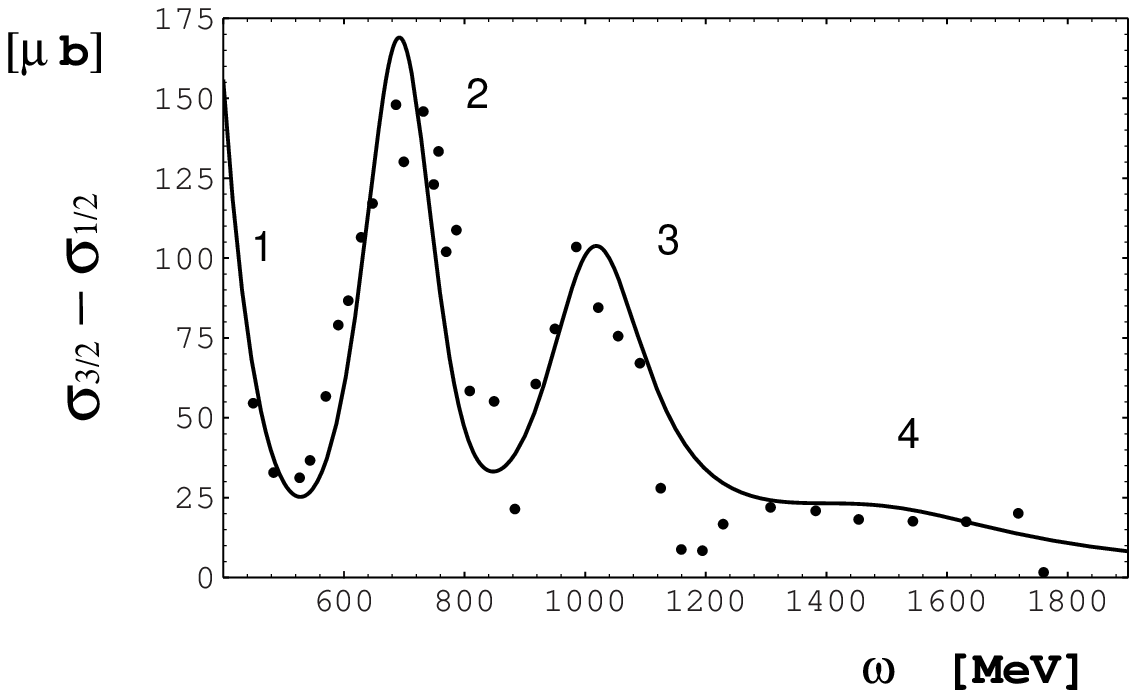}
\caption{Predicted  difference of helicity dependent 
cross sections for the proton
in the second, third and fourth  resonance region compared with experimental
data. The experimental data are taken from \cite{GDH7}. 
The contributions shown in the left panel
 are the tail of the  $P_{33}(1232)$ 
resonance (2), the nonresonant contribution (1),
the  $P_{11}(1440)$, $D_{13}(1520)$ (3),  $S_{11}(1535)$ (4), 
$S_{11}(1650)$,
$D_{15}(1675)$, $F_{15}(1680)$ (5), $D_{33}(1700)$, $F_{35}(1905)$ 
and $F_{37}(1950)$ (6)
resonances. The numbers given in the right panel denote the resonance regions
1--4.
}
\label{ELSA}
\end{figure}
of the peak, 130 MeV $\to$ 120 MeV for the width of the resonance and 
1.26 $\to$ 1.36 for the quantity $A_n$ defined in Eq. (\ref{proportional}).
This means that the integrated cross section and  also the
integrated  energy-weighted cross section remain constant.
For the $D_{13}(1520)$ resonance the shifts were 1520 $\to$ 1490 for 
the position
of the peak, 120 MeV $\to$ 130 for the width of the resonance. All other
parameters remained unmodified compared to the predictions given in Table
\ref{GDHprotonneutron}.

Figure \ref{ELSA} differs from Figure \ref{MAMI} by the fact that the
nonresonant contribution is not eliminated from the figure
but included into the theoretical curve. Indeed, the dip between
the first and the second resonance shown in the right panel
of Figure \ref{ELSA} is strongly influenced by the nonresonant cross sections
represented by curve 1 in the left panel. In the left panel only the stronger
resonances have been identified through a number, though all the relevant
resonances have been shown. As in Figure \ref{MAMI} some shifts of parameters
have been tested  in order to possibly improve on the fit to the experimental
data. The only shift of some relevance was the generation of a small negative 
contribution from the $D_{33}(1700)$ by shifting the quantity $A_n$ from 
$0$ $\to$  $-0.5$.

The importance of the findings made in  Figures \ref{MAMI} and  \ref{ELSA} 
is that  our procedure of  applying  the Walker parameterization
to the helicity dependent cross section difference 
$(\sigma_{3/2}-\sigma_{1/2})$ has been tested and found valid. This makes it
possible to apply this procedure also to other problems as there is
the prediction and interpretation  of the $s$-channel contribution 
to the backward-angle spin-polarizability $\gamma_\pi$.
Furthermore, we have clearly demonstrated that the upper part of the cross
section in the right panel of Figure  \ref{ELSA} is due to the $F_{37}(1950)$
resonance as anticipated  before \cite{GDH7}.

\section{Polarizabilities \label{polarizabilities}}

The appropriate tool for the prediction of electromagnetic polarizabilities 
is to simultaneously apply
the forward-angle sum rule for $(\alpha+\beta)$ and the backward-angle sum
rule for $(\alpha-\beta)$. This leads to the following relations
\cite{schumacher07a} : 
\begin{eqnarray}
&&\alpha=\alpha^s+\alpha^t, \label{pol1}\\
&&\alpha^s=\frac{1}{2\pi^2}\int^\infty_{\omega_0}\left[A(\omega)\sigma(\omega,E1,M2,\cdots)+B(\omega)\sigma(\omega,M1,E2,\cdots)\right]\frac{d\omega}{\omega^2},\label{pol2}\\
&&\alpha^t=\frac12(\alpha-\beta)^t
\end{eqnarray}
and
\begin{eqnarray}
&&\beta=\beta^s+\beta^t,\label{pol3}\\
&&\beta^s=\frac{1}{2\pi^2}\int^\infty_{\omega_0}\left[A(\omega)
\sigma(\omega,M1,E2,\cdots)+B(\omega)\sigma(\omega,E1,M2,\cdots)\right]
\frac{d\omega}{\omega^2},\label{pol4}\\
&&\beta^t=-\frac12(\alpha-\beta)^t,
\end{eqnarray}
with
\begin{eqnarray}
&&\omega_0=m_\pi+\frac{m^2_\pi}{2 m},\label{pol5}\\
&&A(\omega)=\frac12 \left( 1+\sqrt{1+\frac{2\omega}{m}} \right),\label{pol6}\\
&&B(\omega)=\frac12 \left( 1-\sqrt{1+\frac{2\omega}{m}} \right),\label{pol7}\\
&&(\alpha-\beta)^t=\frac{g_{\sigma NN}{\cal M}(\sigma\to\gamma\gamma)}{2\pi m^2_\sigma}
+\frac{g_{f_0 NN}{\cal M}(f_0 \to\gamma\gamma)}{2\pi m^2_{f_0}}
+\frac{g_{a_0 NN}{\cal M}(a_0\to\gamma\gamma)}{2\pi m^2_{a_0}}\tau_3.
\label{pol8}
\end{eqnarray}
In (\ref{pol1}) to (\ref{pol8}) $\omega$ is the photon energy in the
lab. system, $m_\pi$
the pion mass and $m$ the nucleon mass. The quantities $\alpha^s,\beta^s$
are the $s$-channel electric and magnetic polarizabilities, and 
$\alpha^t,\beta^t$ the $t$-channel electric and magnetic polarizabilities,
respectively. The multipole content of the photo-absorption cross-section
enters through
\begin{eqnarray}
&&\sigma(\omega,E1,M2,\cdots)=\sigma(\omega,E1)+ \sigma(\omega,M2)+\cdots,
\label{pol9}\\
&&\sigma(\omega,M1,E2,\cdots)=\sigma(\omega,M1)+ \sigma(\omega,E2)+\cdots,
\label{pol20}
\end{eqnarray}
{\it i.e.} through the sums of cross-sections with change and without 
change of parity during the electromagnetic transition, respectively. The
multipoles belonging to parity change are favored for the
electric polarizability $\alpha^s$ whereas the multipoles belonging to parity
nonchange are favored for the magnetic polarizability $\beta^s$. For the
$t$-channel parts we use the pole representations described in
\cite{schumacher07b}. 

The backward spin-polarizability is given by \cite{schumacher07b,lvov99}
\begin{eqnarray} 
&&\gamma_\pi=\int^\infty_{\omega_0}\sqrt{1+\frac{2\omega}{m}}\left(1+\frac{\omega}{m}
\right)\times \sum_n P_n[\sigma^n_{3/2}(\omega)-\sigma^n_{1/2}(\omega)]
\frac{d\omega}{4\pi^2\omega^3}+\gamma^t_\pi, \label{pol21}\\
&&\gamma^t_\pi=\frac{1}{2\pi m}\left[\frac{g_{\pi NN}
{\cal M}(\pi^0\to\gamma\gamma)}
{ m^2_{\pi^0}}\tau_3
+\frac{g_{\eta NN}{\cal M}(\eta \to\gamma\gamma)}{ m^2_{\eta}}
+\frac{g_{\eta' NN}{\cal M}(\eta'\to\gamma\gamma)}{
  m^2_{\eta'}}\right].\label{pol22}
\end{eqnarray}
where the parity factor is $P_n(E1,M2,\cdots)=-1$ and $P_n(M1,E2,\cdots)=+1$.
The quantities $g_{MNN}$ are the meson-nucleon coupling constants and 
${\cal M}(M\to\gamma\gamma)$ the decay matrix elements.

\subsection{The $t$-channel contributions of  the polarizabilities}

As stated before a very large part of the polarizabilities is not due to 
excitations of the quark-structure of the nucleon but due to the coupling
of the constituent  quarks to scalar and pseudoscalar mesons being 
capable of a two-photon decay. The $\pi^0$ meson pole has been proposed
in 1958 \cite{low58}, the scalar-isoscalar $t$-channel in 1962 \cite{hearn62}.
In several papers  it has been attempted to construct $(\alpha-\beta)^t$
from the reactions $\gamma\gamma\to\pi\pi$ and $\pi\pi\to N\bar{N}$ (see
\cite{schumacher05} for a summary).   This latter approach has been
reconsidered very recently \cite{bernabeu08} in an attempt  to derive
the two-photon width of the $\sigma$ meson from the experimental 
$(\alpha-\beta)^t$.

The first attempts to calculate  $(\alpha-\beta)^t$ from a 
$\sigma$--pole ansatz
has been made in  \cite{lvov97,schumacher05} and presented
in its final form in \cite{schumacher06}. The basic idea of this approach
is that scalar and pseudoscalar mesons showing up as $t$-channel
exchanges  correspond to poles 
 located in the unphysical region of the $s$-$t$ plane and, therefore,
may have properties different from those of  the corresponding 
on-shell particles, except for their two-photon widths $\Gamma_{\gamma\gamma}$.
Furthermore, it may be allowed to represent the $t$-channel exchanges 
in terms of a $q\bar{q}$ core of the meson. For the $\sigma$ meson this 
approach follows the
theory  of Delbourgo and Scadron \cite{delbourgo95} which treats chiral
symmetry breaking in the dynamical linear $\sigma$ model on the quark level.
The application to the present problem is described in 
\cite{schumacher06,schumacher07a,schumacher07b,schumacher08}. The 
results derived in \cite{schumacher07b} are listed in Table 
\ref{polresults2}.

\

\subsection{Numerical results}

The numerical results obtained for the polarizabilities are shown in Tables
\ref{polresults1} and \ref{polresults2}. Table \ref{polresults1} shows
the $s$-channel contributions of the polarizabilities. For the
spin-polarizabilities of resonant states
the scaling factor (\ref{proportionalityconstant}) depending on the
multipolarity mixing has to be taken into account. This is important 
for the $P_{33}(1232)$ resonance where two options {\it viz.} 
$A_n(P_{33}(1232)) = 1.26$ and $A_n(P_{33}(1232)) = 1.0$ are compared with each
other. The corresponding results are given in lines 2 and 3 of
Table \ref{polresults1}. In Table \ref{polresults2} the $t$-channel
contributions are added and a comparison of the results 
with experimental data is
carried out. In addition to the present results also the predictions 
of L'vov and Nathan  \cite{lvov99} are shown.

\begin{table}[h]
\caption{Resonant (lines 2--13) and single-pion nonresonant (lines 16--18)
components of the polarizabilities. The electric and magnetic
  polarizabilities are in unite of $10^{-4}$fm$^3$, the spin
  polarizabilities  in unite of $10^{-4}$fm$^4$.
(a) Present predictions with 
$A_n(P_{33}(1232))=1.26$ (see (\ref{proportionalityconstant})), 
(b) with $A_n(P_{33}(1232))=1.0$. res.+1$\pi$-nr: resonant +1$\pi$ nonresonant
contribution.
}
\begin{center}
\begin{tabular}{l|ll|ll|ll}
\hline
&$\alpha_p$&$\beta_p$&$\alpha_n$&$\beta_n$&$\gamma^{(p)}_\pi$&$\gamma^{(n)}_\pi
$\\ 
\hline
$P_{33}(1232)$&$-1.07$&$+8.32$&$-1.07$&$+8.32$&$+5.11$(a)&$+5.11$(a)\\
&&&&&                                          $+4.05$(b)&$+4.05$(b)\\
$P_{11}(1440)$&$-0.02$&$+0.14$&$-0.01$&$+0.05$&$-0.10$&$-0.04$\\
$D_{13}(1520)$&$+0.68$&$-0.16$&$+0.55$&$-0.13$&$-0.39$&$-0.20$\\
$S_{11}(1535)$&$+0.21$&$-0.05$&$+0.06$&$-0.01$&$+0.13$&$+0.04$\\
$S_{11}(1650)$&$+0.05$&$-0.01$&$+0.00$&$-0.00$&$+0.03$&$+0.00$\\
$D_{15}(1675)$&$+0.01$&$-0.00$&$+0.08$&$-0.02$&$-0.00$&$-0.01$\\
$F_{15}(1680)$&$-0.07$&$+0.25$&$-0.01$&$+0.03$&$+0.13$&$+0.01$\\
$D_{33}(1700)$&$+0.25$&$-0.07$&$+0.25$&$-0.07$&$-0.00$&$-0.00$\\
$F_{35}(1905)$&$-0.01$&$+0.02$&$-0.01$&$+0.02$&$+0.01$&$+0.01$\\
$P_{33}(1920)$&$-0.01$&$+0.02$&$-0.01$&$+0.02$&$+0.01$&$+0.01$\\
$F_{37}(1950)$&$-0.03$&$+0.10$&$-0.03$&$+0.10$&$+0.01$&$+0.01$\\
\hline
sum-res&$-0.01$&$+8.56$&$-0.20$&$+8.31$&$+4.94$(a)&$+4.94$(a)\\
&&&&&                                   $+3.88$(b)&$+3.88$(b)\\
\hline
$E_{0+}$&$+3.19$&$-0.34$&$+4.07$&$-0.43$&$+3.75$&$+4.81$\\
$M^{(3/2)}_{1-}$&$-0.04$&$+0.35$&$-0.04$&$+0.35$&$-0.18$&$-0.18$\\
$(M,E)^{(1/2)}_{1+}$&$-0.06$&$+0.47$&$-0.21$&$+1.44$&$+0.24$&$+0.66$\\
\hline
res.+1$\pi$-nr&$+3.08$&$+9.04$&$+3.62$&$+9.67$&$+8.75$(a)&$+10.23$(a)\\
&&&&&                                          $+7.69$(b)&$+9.17$(b)\\
\hline
\hline
\end{tabular}
\label{polresults1}
\end{center}
\end{table}
\begin{table}[h]
\caption{Lines 2--5: Resonant + 1$\pi$ nonresonant components of
  polarizabilities. 
(a) Present prediction with $A_n(P_{33}(1232))=1.26$, (b) with 
$A_n(P_{33}(1232))=1.0$,
(c) L'vov \cite{lvov99} based on the SAID parameterization of CGLN
amplitudes (d) L'vov \cite{lvov99} based on the HDT parameterization of CGLN
amplitude. Lines 6--8: Scalar and pseudoscalar $t$-channel
components as derived in \cite{schumacher07b}.
Line 9: Nonresonant 2$\pi$ channel with 
 $\gamma N\to \pi\Delta$ as the main component.
Lines 10--13: Total predicted polarizabilities. Lines 14--15:
The experimental results labeled ``exp.'' are those of the analysis
in Ref. \cite{schumacher05},  the experimental results labeled ``exp. corr''
are those of the analysis in Ref. \cite{schumacher07b}. }
\begin{center}
\begin{tabular}{l|ll|ll|ll}
\hline
&$\alpha_p$&$\beta_p$&$\alpha_n$&$\beta_n$&$\gamma^{(p)}_\pi$&
$\gamma^{(n)}_\pi$\\
\hline
res.+1$\pi$ nr&$+3.08$&$+9.04$&$+3.62$&$+9.67$&$+8.75(a)$&$+10.23$(a)\\
&&&&&                                          $+7.69(b)$&$+9.17$(b)\\
&&&&&                                          $+7.31$(c)&$+9.35$(c)\\
&&&&&                                          $+8.37$(d)&$+9.76$(d)\\
$\sigma/\pi^0$-$t$-ch.&$+7.6$&$-7.6$&$+7.6$&$-7.6$&$-46.7$&$+46.7$\\
$f_0/\eta$-$t$-ch.&$+0.3$&$-0.3$&$+0.3$&$-0.3$&$+1.2$&$+1.2$ \\
$a_0/\eta'$-$t$-ch.&$-0.4$&$+0.4$&$+0.4$&$-0.4$&$+0.4$&$+0.4$  \\
$\gamma N\to \pi \Delta$&$+1.4$&$+0.4$&$+1.5$&$+0.4$&$-0.28$&$-0.23$\\
\hline
sum&$12.0$&$+1.9$&$+13.4$&$+1.8$&$-36.6$(a)&$+58.3$(a)\\
&&&&&                            $-37.7$(b)&$+57.2$(b)\\
&&&&&                            $-38.1$(c)&$+57.4$(c)\\
&&&&&                            $-37.0$(d)&$+57.8$(d)\\
\hline
exp.&$+12.0\pm 0.6$&$+1.9\mp 0.6$&$+12.5\pm 1.7$&$+2.7\mp1.8$&$-38.7\pm 1.8$
&$58.6\pm 4.0$\\
exp. corr.&$+12.0\pm 0.6$&$+1.9\mp 0.6$&$+13.4\pm 1.0$&$+1.8\mp1.0$&$-38.7\pm 1.8$
&$57.6\pm 1.8$\\
\hline
\end{tabular}
\label{polresults2}
\end{center}
\end{table}
Some remarks have to be made concerning the contribution labeled 
$\gamma N\to \pi\Delta$. This contribution contains all reactions with more
than one pion in the final state where the $\gamma N\to \pi\Delta$
contribution is the most prominent one, except for those where the nucleon
resonant states are the intermediate states. For the spin polarizabilities
the predictions are taken from \cite{lvov99}. For the electric and magnetic
polarizabilities the multipolarity ratio $E1/M1$ has to be 
known which was 
adopted  to be $70\%(E1)$ and $30\%(M1)$ for sake of 
consistency of the 
prediction with the experimental data. 
At present there is no easy possibility
to determine the multipolarity ratio $E1/M1$ 
from theoretical arguments only.

For the $s$-channel parts of the spin-polarizabilities four different 
predictions are compared with each other
in lines 2 -- 5 of Table \ref{polresults2}.
We notice that  the present
results (a) are in better agreement with the L'vov result \cite{lvov99} (d)
and the present results (b) with the L'vov \cite{lvov99}
results (c). 
The deviations of the predictions (a) -- (d) in lines 10 --13 from the 
experimental results
are  smaller than the experimental errors. Therefore,
improvements may be  expected from higher precision of the experimental data
in the  first place.

\section{Discussion \label{discussion} }

In the present investigation we have shown that the level scheme of the nucleon
corresponding to the first three oscillator shells of the $SU(6)$-HO model
is  well understood. As a specific feature we have found that for I=1/2 the
photo-excitation strengths of resonant states in the neutron are smaller
than the corresponding quantities of  the proton. An explanation for this
specific feature is provided by the quark model where this feature shows
up as a consequence of the different electric charges of the quarks 
entering into electromagnetic transition matrix elements. One consequence
of this feature is that the radiative width of the $F_{15}(1680)$ resonance 
is only exceptionally  large in the proton but rather small in the neutron.
Therefore, for the neutron the third resonance region showing up 
in the
photoexcitation spectrum is not dominated by the $F_{15}(1680)$ resonance
but by the $D_{33}(1700)$ resonance. An exception of this general intensity
rule  is only observed for the $D_{15}(1675)$ resonance where the Moorhouse
\cite{moorhouse66}  selection rule predicts a very small photoexcitation
strength for the proton.

A further success of the present approach is the quantitative
representation  of  helicity dependent cross section
$(\sigma_{3/2}-\sigma_{1/2})$ in terms of resonant and nonresonant
contributions and the good agreement of the predictions with experimental
data.  An interesting result obtained in this connection is 
that the resonant structure of the $(\sigma_{3/2}-\sigma_{1/2})$ cross 
section is much more dependent on the mixture of multipoles than the resonant
structure of the total cross section $\sigma_T$. For the $P_{33}(1232)$
resonance the $E_{1+}$ component enhances the partial 
$(\sigma_{3/2}-\sigma_{1/2})$ cross section by $ 26\%$.
The experimental spectra shown in Figures \ref{MAMI}
and \ref{ELSA} are well understood. Especially the peak structure
in Figure \ref{ELSA} around 1500 MeV can clearly be interpreted in terms
of the $F_{37}(1950)$ resonance. 

For the polarizabilities $\alpha$, $\beta$ and $\gamma_\pi$ the $P_{33}(1232)$
resonance dominates the resonant $s$-channel contribution  and the 
electric-dipole
``pion-cloud'' amplitude  $E_{0+}$ the nonresonant $s$-channel contribution.
In addition the $\sigma$-meson $t$-channel exchange makes a large contribution
to the electric ($\alpha$) and magnetic ($\beta$) polarizabilities. For the
electric polarizability  the $\sigma$-meson $t$-channel contribution is about
twice as large as the ``pion-cloud'' $E_{0+}$ contribution. For the magnetic
polarizability the  $\sigma$-meson $t$-channel contribution explains the
pronounced diamagnetism of the nucleon. For the backward spin-polarizability
the $\pi^0$ $t$-channel contribution  is larger than the $s$-channel
contributions by a factor of $\approx 5$. 

It should be noted that a comparison between results from chiral perturbation
theory and dispersion theory has been carried out in \cite{schumacher07a}.
\clearpage
\newpage
\appendix

\section{Level scheme of the nucleon} 
\begin{figure}[h]
\begin{minipage}[b]{50mm}
\includegraphics[width=1.0\linewidth]{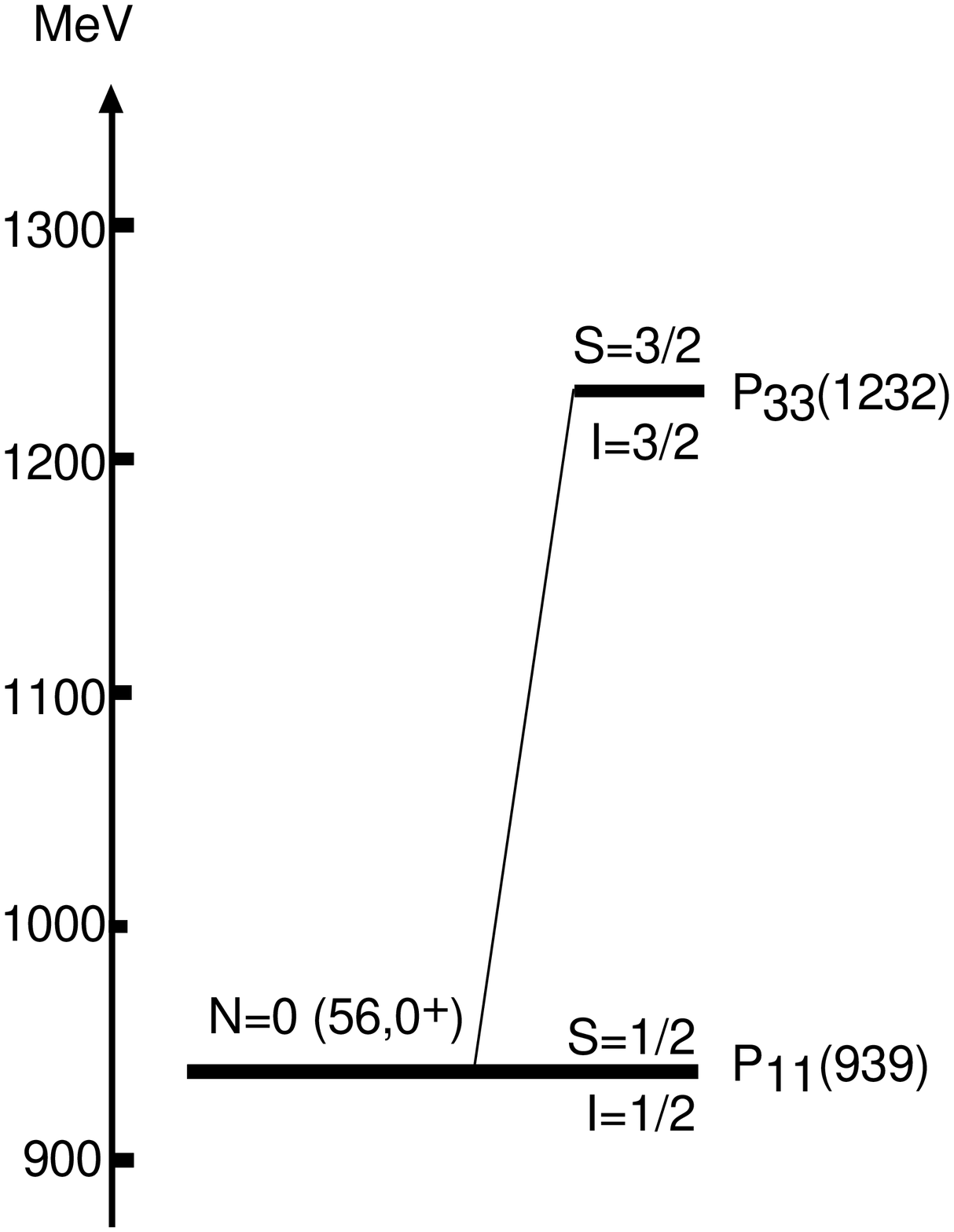}
\end{minipage}
\begin{minipage}[b]{50mm}
\includegraphics[width=1.0\linewidth]{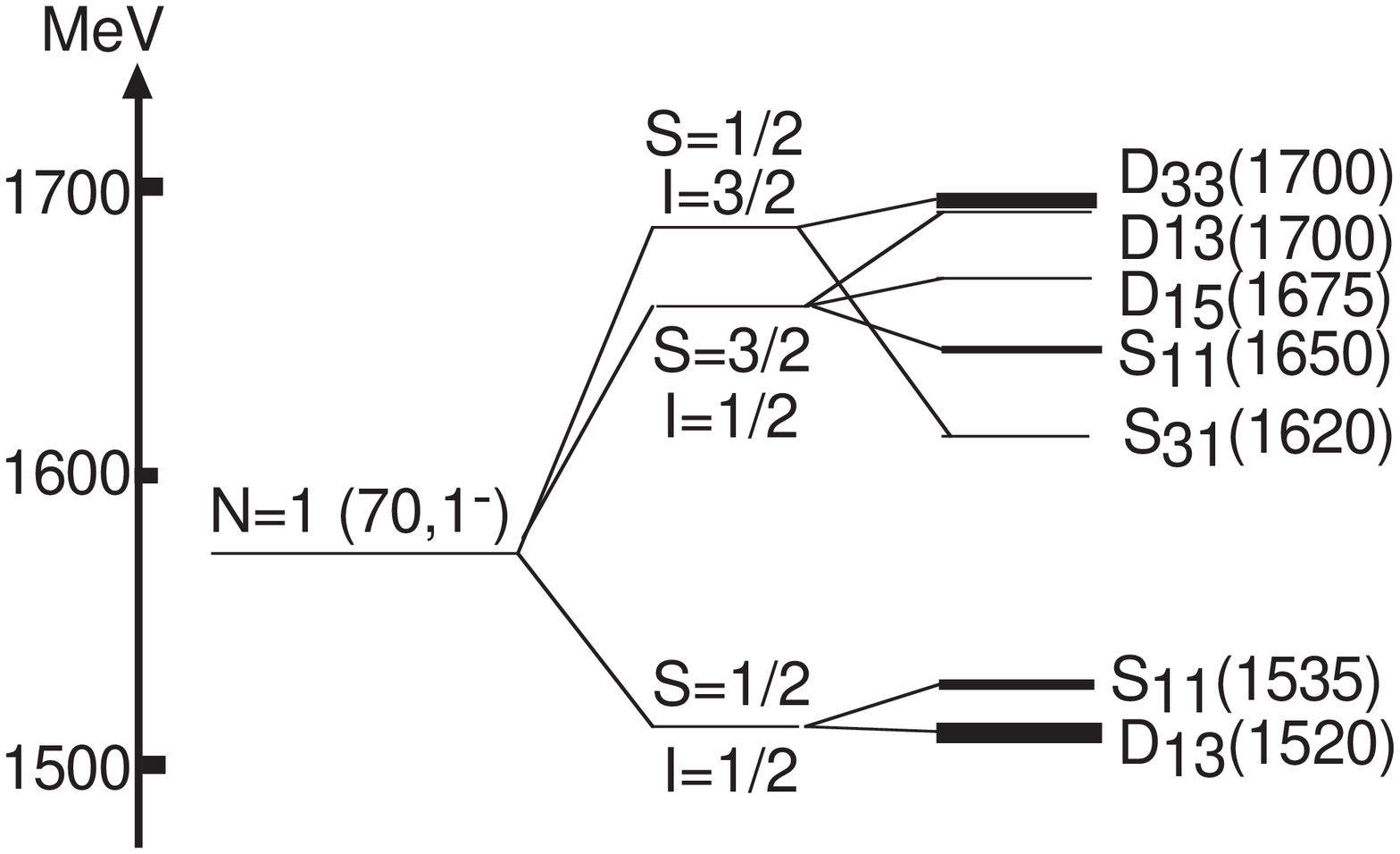}
\vspace{65mm}
\end{minipage}
\begin{minipage}[b]{50mm}
\includegraphics[width=1.00\linewidth]{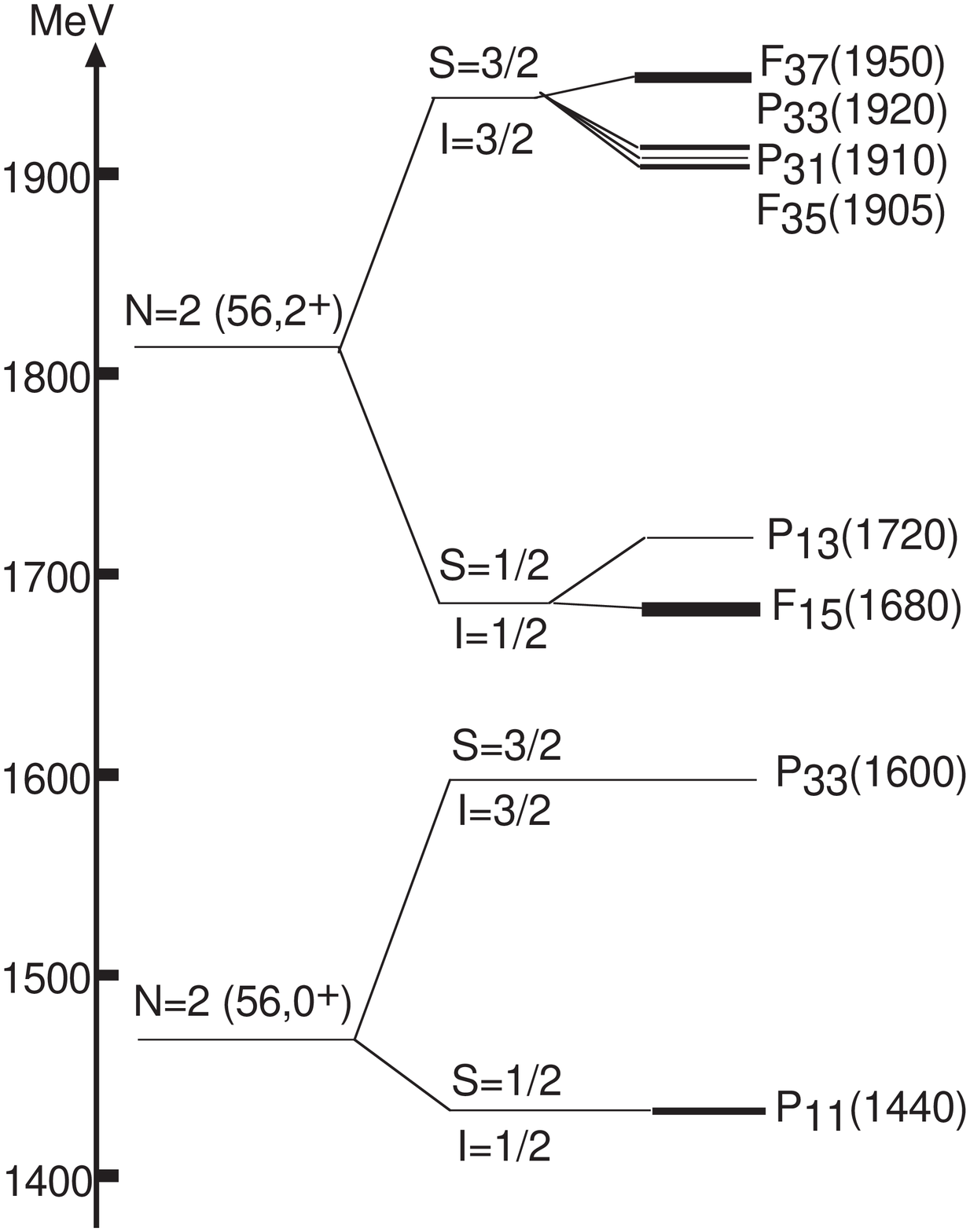}
\vspace{56mm}
\end{minipage}
\caption{The $N=0$ $(56,0^+)$ (left), $N=1$ 
$(70,1^-)$ (middle), and $N=2$  $(56,0^+)$, $(56,2^+)$ (right)
partial level schemes of the nucleon} 
\label{secondshell}
\end{figure}

Figure \ref{secondshell} shows the level scheme of the nucleon
corresponding to the first three HO shells. 
In total four resonance regions are formed. The first resonance region
is solely due to the $P_{33}(1232)$ state  which as a $I=3/2$ state
is of equal photon-excitation
strength in the proton and the neutron. The second resonance
region is due to the $D_{13}(1520)$ and the $S_{11}(1535)$ states. As $I=1/2$
states the strength is larger in the proton than in the neutron in both
resonances. The third resonance region consists of the $F_{15}(1680)$
and the $D_{33}(1700)$ states. For  the proton the two contributions are of
approximately equal strength,  whereas for the neutron
according to the general rule for
$I=1/2$ states the $F_{15}(1680)$ contribution is largely suppressed. 
The fourth
resonance is due to the $F_{37}(1950)$ state which is of equal strength in the
proton and the neutron.
\clearpage
\newpage

\section{Resonance parameters of the nucleon}

\begin{table}[h]
\caption{Resonance parameters of proton  resonant states for the first three
oscillator shells.
The sixth column denoted by $I_{int.}$(p) contains the integrated 
photo-absorption
cross sections for the proton as following from the data given in columns
2--5. The ordering of levels is the same as in Table
\ref{singleparticletable-1}.
}
\vspace{3mm}
\setlength{\extrarowheight}{5pt}
\begin{center}
\begin{tabular}{lccccc}
\hline
& $10^3\,A_{1/2}$ & $10^3\,A_{3/2}$ & $\Gamma_r$ & $I_r$&$I_{int.}$(p)\\
&$[{\rm GeV}^{-1/2}]$& $[{\rm GeV}^{-1/2}]$& [MeV]& [$\mu$b]& [$10^3\,\mu$b MeV]\\ 
\hline
$P_{33}(1232)$ &$-139 \pm 3$ &$ -257\pm 4$ & $130\pm 5$ & $390\pm 10$&$80.0\pm
4.0$ \\
$P_{11}(1440)$& $-65\pm 4$ & $-$ & $350\pm 70$ & $6.1 \pm 1.4$&$3.4\pm 0.8$\\
$P_{33}(1600)$& $-23\pm 20$ & $-9\pm 21$ & $330\pm 50$ & $0.8 \pm 0.8$&$
0.7\pm 0.6$\\
$S_{11}(1535)$& $+90\pm 20$ & $-       $ & $150\pm 30$ & $26  \pm 10 $&
$8.8\pm 3.5$\\
$D_{13}(1520)$& $-24\pm 9 $ & $+166 \pm 5$& $120\pm 10$ & $113 \pm 11 $&$
34.6\pm 3.5$\\
$S_{31}(1620)$& $+27\pm 11$ & $- $&          $140\pm 20$ & $2.4 \pm 1.4$&
$0.9\pm 0.5$\\
$D_{33}(1700)$& $+104\pm 15$ & $+85\pm 22 $& $270\pm 70$ & $29 \pm 12$&
$18.0\pm 3.6$\\
$S_{11}(1650)$& $+53\pm 16$ & $-$& $160\pm 10$ & $7.8 \pm 2.3$&
$3.1\pm 0.9$\\
$D_{13}(1700)$& $-18\pm 13$ & $-2\pm 24$& $100\pm 20$ & $1.4 \pm 1.0$&
$0.4\pm 0.3$\\
$D_{15}(1675)$& $+19\pm 8$ & $+15\pm 9$& $150\pm 10$ & $1.7 \pm 1.0$&
$0.7\pm 0.4$\\
$P_{13}(1720)$& $+18\pm 30$ & $-19\pm 20$& $150\pm 30$ & $1.9 \pm
1.9$&$0.8\pm 0.8$\\
$F_{15}(1680)$& $-15\pm 6$ & $+133\pm 12$& $130\pm 10$ & $60 \pm 12$&
$19.9\pm 2.0$\\
$P_{31}(1910)$& $+3\pm 14$ & $-$& $240\pm 20$ & $0.01 \pm 0.01$&$< 0.1$\\
$P_{33}(1920)$& $+40\pm 14$ & $+23\pm 17$& $200\pm 40$ & $4.0 \pm 2.5$&
$2.3\pm 1.6$\\
$F_{35}(1905)$& $+26\pm 11$ & $-45\pm 20$& $300\pm 60$ & $3.5 \pm 3.0$&
$2.7\pm 1.5$\\
$F_{37}(1950)$& $-76\pm 12$ & $-97\pm 10$& $280\pm 20$ & $20.3 \pm 5.0$&
$15.8\pm 4.0$\\
\hline
\end{tabular}
\end{center}
\label{resonanceparametersproton}
\end{table}

\begin{table}[h]
\caption{Resonance parameters of neutron resonant states for the first three
oscillator shells. The sixth column denoted by $I_{int.}$(n)
contains the integrated photo-absorption cross sections for the neutron as
following from the data given in columns 2-5. The ordering of levels is 
the same as in Table
\ref{singleparticletable-1}.}
\vspace{3mm}
\setlength{\extrarowheight}{5pt}
\begin{center}
\begin{tabular}{lccccc}
\hline
& $10^3\,A_{1/2}$ & $10^3\,A_{3/2}$ & $\Gamma_r$ & $I_r$ &$I_{int.}$(n)\\
&$[{\rm GeV}^{-1/2}]$& $[{\rm GeV}^{-1/2}]$& [MeV]& [$\mu$b]  & [$10^3\,\mu$b MeV]\\
\hline
$P_{33}(1232)$ &$-139 \pm 3$ &$ -257\pm 4$ & $130\pm 5$ & $390\pm 10$&
$80.0\pm 4.0$\\
$P_{11}(1440)$& $+40\pm 10$ & $-$ & $350\pm 70$ & $2.3\pm 1.2$&
$1.3\pm 0.7$\\
$P_{33}(1600)$& $-23\pm 20$ & $-9\pm 21$ & $330\pm 50$ & $0.8 \pm 0.8$&
$0.7\pm 0.6$\\
$S_{11}(1535)$& $-46\pm 27$ & $-       $ & $150\pm 30$ &$7\pm 4$&
$2.4\pm1.4$ \\
$D_{13}(1520)$& $-59\pm 9 $ & $-139 \pm 11$& $120\pm 10$ &$ 91\pm 15$&
$27.9\pm 4.6$\\
$S_{31}(1620)$& $+27\pm 11$ & $- $&          $140\pm 20$ & $2.4 \pm 1.4$&
$0.9\pm 0.5$\\
$D_{33}(1700)$& $+104\pm 15$ & $+85\pm 22 $& $270\pm 70$ & $29 \pm 12$&
$18.0\pm 3.6$\\
$S_{11}(1650)$& $-15\pm 21$ & $-$& $160\pm 10$ &$0.6\pm 0.6$&
$0.2\pm 0.2$\\
$D_{13}(1700)$& $0\pm 50$ & $-33\pm 44$& $100\pm 20$ &$0.5\pm
0.5$& $0.1\pm 0.1$\\
$D_{15}(1675)$& $-43\pm 12$ & $-58\pm 13$& $150\pm 10$ &$15\pm
5$& $6.2\pm 2.1$\\
$P_{13}(1720)$& $+1\pm 15$ & $-29\pm 61$& $150\pm 30$ &$2.3\pm 3.0$& 
$1.0\pm 1.0$\\
$F_{15}(1680)$& $+29\pm 10$ & $-33\pm 9$& $130\pm 10$ &$6.5\pm
 2.8$&$2.2\pm 0.9$ \\
$P_{31}(1910)$& $+3\pm 14$ & $-$& $240\pm 20$ & $0.01 \pm 0.01$&$<\, 0.1$\\
$P_{33}(1920)$& $+40\pm 14$ & $+23\pm 17$& $200\pm 40$ & $4.0 \pm 2.5$&
$2.3\pm 1.6$\\
$F_{35}(1905)$& $+26\pm 11$ & $-45\pm 20$& $300\pm 60$ & $3.5 \pm 3.0$&
$2.7\pm 1.5$\\
$F_{37}(1950)$& $-76\pm 12$ & $-97\pm 10$& $280\pm 20$ & $20.3 \pm 5.0$&
$15.8\pm 4.0$\\
\hline
\end{tabular}
\end{center}
\label{resonanceparametersneutron}
\end{table}

\end{document}